\documentclass[webpdf,contemporary,large]{oup-authoring-template}%

\usepackage{paralist}

\usepackage{graphicx}  
\usepackage{xcolor}    

\definecolor{codegray}{gray}{0.9}

\usepackage{threeparttable}
\usepackage{booktabs}
\usepackage{tabularx} 
\usepackage{array}    
\usepackage{amsmath,amssymb,amsthm,bm,mathtools}

\usepackage[loadonly]{enumitem}
\usepackage{tikz}
\usetikzlibrary{arrows.meta, positioning,fit, backgrounds,shapes.geometric}
\usepackage{pgfplots}\pgfplotsset{compat=1.16}

\usepackage{tabularx}
\usepackage{tikz}
\usetikzlibrary{arrows.meta,positioning,fit,backgrounds,shapes.geometric}

\theoremstyle{plain}

\usepackage{amsmath}

\newcommand{\pic}{\pi_{c}}                       
\newcommand{\uc}{u_{c}}                          
\newcommand{\Tdir}{T_{\mathrm{dir}}}
\newcommand{\That}{\widehat{T}_{\mathrm{dir}}}
\newcommand{\Tdec}{T_{\mathrm{dec}}}
\newcommand{\kalign}{\omega_{\mathrm{align}}}
\newcommand{\gcurv}{\gamma_{\mathrm{curv}}}
\newcommand{\bmerit}{\bm{\beta}_{\mathrm{merit}}}
\newcommand{\aout}{a_{\mathrm{out}}}
\newcommand{\rhoout}{\rho_{\mathrm{out}}}
\newcommand{\zout}{z_{\mathrm{out}}}
\newcommand{\Ftil}{\widetilde{F}}
\newcommand{\Fbtil}{\widetilde{\bm F}}

\newcommand{\pibar}{\bar{\pi}}
\newcommand{\pitarget}{\bar{\pi}_{\mathrm{target}}}

\newcommand{\Dfixed}{\mathcal{D}_{\mathrm{fixed}}}

\newcommand{\Beff}{B_{\mathrm{eff}}}

\newcommand{\Bcal}{B_{\mathrm{cal}}}

\newcommand{\logit}{\operatorname{logit}}
\newcommand{\sd}{\operatorname{sd}}
\newcommand{\Var}{\operatorname{Var}}
\newcommand{\Cor}{\operatorname{Cor}}
\newcommand{\E}{\operatorname{E}}
\newcommand{\Bern}{\operatorname{Bernoulli}}

\newcommand{\dd}{\delta}

\newcommand{\figslot}[3][\linewidth]{%
	\IfFileExists{#2}{\includegraphics[width=#1]{#2}}{%
		\fbox{\parbox[c][#3][c]{0.95\linewidth}{\centering\small\ttfamily
				Figure file \detokenize{#2} not found.\\
				Place the exported figure at this path to include it.}}}}

\definecolor{cProt}{HTML}{E08214}
\definecolor{cMerit}{HTML}{2C7FB8}
\definecolor{cFair}{HTML}{31A354}
\definecolor{cUnfair}{HTML}{D7301F}
\definecolor{cOracle}{HTML}{1B9E77}
\definecolor{cMerOnly}{HTML}{D95F02}
\definecolor{cSpat}{HTML}{7570B3}

\usepackage{enumitem}

\renewlist{description}{description}{1}
\setlist[description]{style=nextline, labelwidth=!, leftmargin=1.5em}

\usepackage{setspace}
\usepackage{xcolor}

\graphicspath{{figures/}}

\theoremstyle{thmstyleone}%

\newtheorem{lemma}{Lemma}%
\theoremstyle{thmstyletwo}%
\theoremstyle{thmstylethree}%
\newtheorem{definition}{Definition}

\newcommand{\ig}[1]{#1}
\newcommand{\chg}[1]{\textcolor{black}{#1}}
\newcommand{\fix}[1]{\textcolor{black}{#1}}
\newcommand{\fixcolor}{\color{black}}
\newcommand{\BFair}{\textsc{B-Fair}}

\begin{document}
	\journaltitle{The Journal of the Royal Statistical Society, Series A (Statistics in Society)}
	\DOI{DOI added during production}
	\copyrightyear{YEAR}
	\pubyear{YEAR}
	\vol{XX}
	\issue{x}
	\access{Published: Date added during production}
	\appnotes{Paper}
	
	\firstpage{1}
	
	\title[Auditing spatial fairness]{A fairness threshold is not a portable standard: a statistical protocol for auditing algorithmic decisions about people in places}
	
	\author[1]{Gandhi Jafta\ORCID{0000-0003-1889-2185}}
	\author[1]{Inger Fabris-Rotelli\ORCID{0000-0002-2192-4873}}
	\author[2]{Emma Ruttkamp-Bloem\ORCID{0000-0003-0299-6406}}
	\author[1]{Janet van Niekerk\ORCID{ 0000-0002-4334-2057}}
	\author[1]{Iketle Maharela\ORCID{0009-0006-0693-9563}}
	\address[1]{\orgdiv{Department of Statistics}, \orgname{University of Pretoria}}
	\address[2]{\orgdiv{Department of Philosophy}, \orgname{University of Pretoria}}
	
	\corresp[$\ast$]{Gandhi Jafta. \href{jafta.gandhi@up.ac.za}{jafta.gandhi@up.ac.za}}
	
	\received{Date}{0}{Year}
	\revised{Date}{0}{Year}
	\accepted{Date}{0}{Year}
	\abstract{Regulators increasingly require fairness audits of automated decisions, and the Royal Statistical Society asks that the evidence submitted about AI systems be statistically grounded and tested under distribution shift. To our knowledge, no audit framework specifies how an audit should proceed when decisions are made about people in places, where outcomes cluster geographically for reasons that may or may not be legitimate and where location itself carries information about protected status. We propose a six-step protocol for such audits, built on \BFair{}, a Bayesian measure of disparity among merit-comparable individuals that is aligned with protected context. Two steps are declarations that \chg{should} be made before the outcomes are examined: who counts as comparable, and which spatial variation in outcomes a fair system may show. Three are computations: nuisance spatial models fitted without the outcomes, a posterior distribution of the disparity, and a calibration of the decision against the audited population itself. The last is a report that a regulator can read as evidence rather than as a score. A simulation study shows why each step is needed. A fair world that ignores legitimate spatial clustering flags 15.9\% of fair systems. A decision threshold carried over from another population flags up to 21.0\%, because what a fair system looks like depends on the geography audited; calibrating each audit on its own population \ig{brings this back to between 3.4\% and 6.2\%, close to the nominal 5\%, while lowering the feasible audit's detection rate only from 27.4\% to 22.4\%}. \chg{The failure of a borrowed threshold persists across effect sizes, ranges of spatial clustering and sample sizes, and in larger samples it becomes relatively worse, not better.} How much spatial variation is legitimate cannot be learned from the data, so it \chg{should} be declared in advance. A worked audit report shows that the same transported threshold both condemns a fair city and clears an unfair one, while the protocol decides both correctly. A fairness threshold is not a portable standard; an audit procedure can be.}
	\maketitle
	
	\section{Introduction}
	\label{sec:intro}
	Artificial intelligence systems should be Secure, Accurate, Fair, and Explainable (SAFE)\ig{\footnote{Royal Statistical Society (2026), \emph{AI is Statistics: Why statistical thinking is vital for the effective, ethical and safe use of AI}, 2 March 2026, \url{https://rss.org.uk/RSS/media/File-library/Policy/2026/AI-is-Statistics-FINAL.pdf}\label{fn:rss}}}. Fairness audits are how the third of these is meant to be checked, and they are moving from good practice to legal requirement.  The EU Artificial Intelligence Act classifies certain AI systems used to evaluate the creditworthiness or establish the credit score of natural persons as high-risk and requires providers of high-risk systems to apply data-governance practices for identifying, preventing and mitigating possible biases \citep{euaiact2024}. New York City Local Law~144\footnote{\url{https://www.nyc.gov/site/dca/about/automated-employment-decision-tools.page}} prohibits employers and employment agencies from using an automated employment decision tool for hiring or promotion unless it has undergone an independent bias audit within the preceding year, a summary of the audit results has been made publicly available, and the prescribed notices have been given. In South Africa, Section~71 of the Protection of Personal Information Act restricts decisions that have legal consequences for, or substantially affect, a data subject when they are based solely on automated processing intended to profile that person, subject to specified exceptions and safeguards\ig{\footnote{Protection of Personal Information Act 4 of 2013, Republic of South Africa, Government Gazette 37067, \url{https://www.gov.za/documents/protection-personal-information-act}}}. These instruments say that an audit must take place. They say little about how its evidence should be judged. The Royal Statistical Society identifies exactly this gap:
	
	\begin{quote}
		\textit{In regulated sectors, organisations are expected to attest to the safety and effectiveness of the systems they use. Yet evaluation metrics are often inconsistent, opaque, or presented without information about sampling, uncertainty or variation across subgroups. A single accuracy figure, absent measures of uncertainty or robustness, can give a misleading picture of reliability. Regulators, therefore, need statistical capacity to judge whether the evidence presented is meaningful and comparable.}
		\hfill --- \textbf{Royal Statistical Society}, \emph{AI is Statistics}\ig{\footnote{Section~4, ``Implications for policy''; full reference in footnote~\ref{fn:rss}.}}
	\end{quote}
	
	A single fairness score is like the average temperature of a hospital: it tells you everything is fine on average, while half the patients are freezing and the other half have a fever. A multi-dimensional ethical dilemma cannot be squeezed into a one-dimensional math problem.  This paper asks what an audit must contain for a regulator to treat a finding of unfairness as meaningful and comparable evidence, in the setting where the question is hardest: decisions about people in places. Credit, insurance, policing and public services are all allocated across a geography in which location carries information about race, class and history. Redlining, the restriction of financial and other services in particular neighbourhoods according to their racial or ethnic composition, showed how an apparently geographic decision rule can encode protected disadvantage \citep{sacharidis2023auditing,saxena2025legally,saxena2024spatial}.
	
	Spatial data break the usual audit in two ways, and both are failures of the evidential standard rather than of the fairness measure. The first concerns the fair world. Outcomes may cluster in space for legitimate reasons, such as local labour markets or local risks, and in any one city such clustering co-varies with the protected group's geography by chance. An audit whose model of a fair system has no room for legitimate clustering mistakes that chance alignment for discrimination. Which spatial variation is legitimate is a normative question that the outcome data cannot settle, so it must be declared. This ought not to discourage statisticians because this is an inherently multidisciplinary  pursuit \ig{\citep{jafta2021evaluation}}. Further\ig{,} these normative choices must be made with the relevant stakeholders. For example urban sociologists and geographers can help identify historical migration, zoning laws, and economic clustering that explain why a demographic group lives near a specific risk or market.
	
	The second concerns the unfairness threshold. Even in a perfectly fair system, chance produces some comparable pairs in which the person in the more protected area receives the worse outcome, so a fair system does not score zero. How far above zero it scores depends on the realised geography, and so a threshold calibrated in one population does not transfer to another. A regulator who sets a single number risks condemning fair systems in one city and clearing unfair ones in the next.
	
	To our knowledge, no audit framework specifies how spatial dependence should enter the fair world against which a system is judged, or how an audit's threshold should be set when the fair-world benchmark varies with geography. Our contribution is such a framework. We propose a six-step protocol for spatial fairness audits in Section~\ref{sec:protocol} that separates the normative declarations a regulator or stakeholder must make from the computations an auditor performs, and that ends in a report a regulator can read. The protocol is built on \BFair{} \citep{jafta2026bfair}, a Bayesian measure of protected-context-aligned disparity among merit-comparable individuals, which we summarise in Section~\ref{sec:bfair}; the statistical properties of its decision layer are developed in a companion thesis \citep{jafta2026calibrating}. Section~\ref{sec:evidence} uses simulated cities, in which the truth is known, to show what goes wrong when each step is omitted. Section~\ref{sec:worked} presents a worked audit report. Section~\ref{sec:regulators} sets out what a regulator should require, and Section~\ref{sec:limits} states the limits of what such an audit can certify.
	
	\section{\fix{Background and related work}}
	\label{sec:litreview}
	
	Algorithmic fairness research has developed largely along two lines: group fairness and individual fairness. Group fairness enforces statistical parity conditions across protected groups, such as demographic parity or equal opportunity \citep{feldman2015certifying,hardt2016equality}. In contrast, individual fairness requires that similar individuals receive similar treatment, independent of group membership \citep{dwork2012fairness}. Both traditions remain important as starting points for the field, but neither is our sole focus here. This paper is concerned instead with a problem that cuts across both: spatial proxy bias, the failure of standard fairness metrics to account for the spatial structure of the populations they audit and uncertainty quantification. 
	
	Geographic location often encodes structural and historical processes and is rarely \textit{neutral}  \citep{dadashpoor2025defining}. Omitting location from a fairness audit risks allowing algorithms to encode and reinforce these historical patterns that discriminate based on protected attributes \citep{baresi2025neo,saxena2025legally}.
	
	Recent work has begun to address this directly by proposing mechanisms for spatial data fairness \citep{shaham2022models} and \ig{fair spatial-indexing (k-d tree) partitions of space that reduce miscalibration across regions \citep{shaham2024fairkd}}. \ig{\citet{xie2022fairness} and \citet{he2022sailing}} show that \ig{location-based fairness measures are sensitive to how space is partitioned, and that this sensitivity can be exploited to steer a model towards a chosen fairness outcome}, exposing the fragility of fairness metrics to the long-studied Modifiable Areal Unit Problem (MAUP) \citep{openshaw1983modifiable}. \citet{sacharidis2023auditing} develops an auditing framework that scans geographic space to identify regions whose outcome distribution differs significantly \ig{inside and outside the region}.
	
	Standard fairness metrics such as demographic parity \citep{dwork2012fairness}, equal opportunity and equalised odds \citep{hardt2016equality} are often reported as point estimates, with limited attention to sampling variation, model uncertainty, or robustness across sparse protected populations \citep{Giudici09072026}. Fairness assessments can be unstable when effective group sizes are small \citep{barrainkua2024uncertainty, Giudici09072026}, \ig{and standard resampling procedures can fail when outcomes are dependent \citep{kunsch1989jackknife,lahiri2003resampling}}. Bayesian formulations provide a route to propagating model uncertainty \citep{barrainkua2024uncertainty}, but posterior uncertainty alone does not establish repeated-design calibration\ig{ \citep{rubin1984bayesianly}}. \BFair{} therefore combines posterior propagation with explicit simulation-based evaluation of the decision rule. 
	
	Hypothesis testing for fairness is not itself new. \citet{cherian2024statistical} frame fairness auditing as multiple hypothesis testing and use the bootstrap to bound performance disparities simultaneously over collections of groups, building on single-test formulations such as \citet{diciccio2020evaluating}. In the spatial setting, \citet{sacharidis2023auditing} define spatial fairness as statistical independence of outcomes from location and test it against regional alternatives. These approaches calibrate against a null of group parity or of no spatial variation in outcomes. Our setting differs in two respects: the estimand is individual and directional, and the fair null must permit legitimate spatial dependence in outcomes. A null of spatial independence would treat the legitimate clustering of our principal stress scenario as evidence against fairness, so that its rejection rate would measure the strength of benign clustering rather than unfairness.
	
	Two neighbouring literatures are closest to ours. \citet{maity2021statistical} develop hypothesis tests for individual fairness with asymptotic control of the Type~I error, for a fixed model and a fair metric on inputs; Bayesian approaches \ig{account for model uncertainty in fair decision rules \citep{dimitrakakis2019bayesian} and} estimate group-fairness metrics with posterior uncertainty \citep{foulds2020bayesian,ji2020can}. Neither addresses spatial dependence in outcomes or the dependence of the fair null on the audited population.
	
	The label \emph{individual} requires a qualification. In the sense of \citet{dwork2012fairness}, individual fairness is a Lipschitz condition on the treatment of similar individuals that makes no reference to group membership\ig{: for a decision map $M$, a distance $d$ between individuals and a distance $D$ between treatments, $D\{M(x),M(y)\}\le L\,d(x,y)$ for all individuals $x$ and $y$, so that two individuals can be treated differently only in proportion to how different they are}. \BFair{} is individual in its unit of comparison, since it compares pairs of individuals who are close in residualised merit, but it weights each pair by $|\pi(s_i)-\pi(s_j)|$, so it measures the part of individual-level inconsistency that is aligned with protected context. It is therefore a hybrid: individual in construction and group-aligned in what it counts. An outcome difference between comparable peers in the same protected context contributes nothing in the magnitude of the estimand, however large it is. These differences are the subject of the Lipschitz \ig{condition} and lie outside the present estimand.
	
	Existing work motivates both spatially aware fairness assessment and uncertainty quantification. The unresolved problem addressed here is an inferential one: how should a continuous, spatially structured individual fairness statistic be calibrated against plausible fair-world variation? \fix{Section~\ref{sec:protocol} answers with a procedure rather than a statistic.}

	\section{\fix{A six-step protocol for spatial fairness audits}}
	\label{sec:protocol}
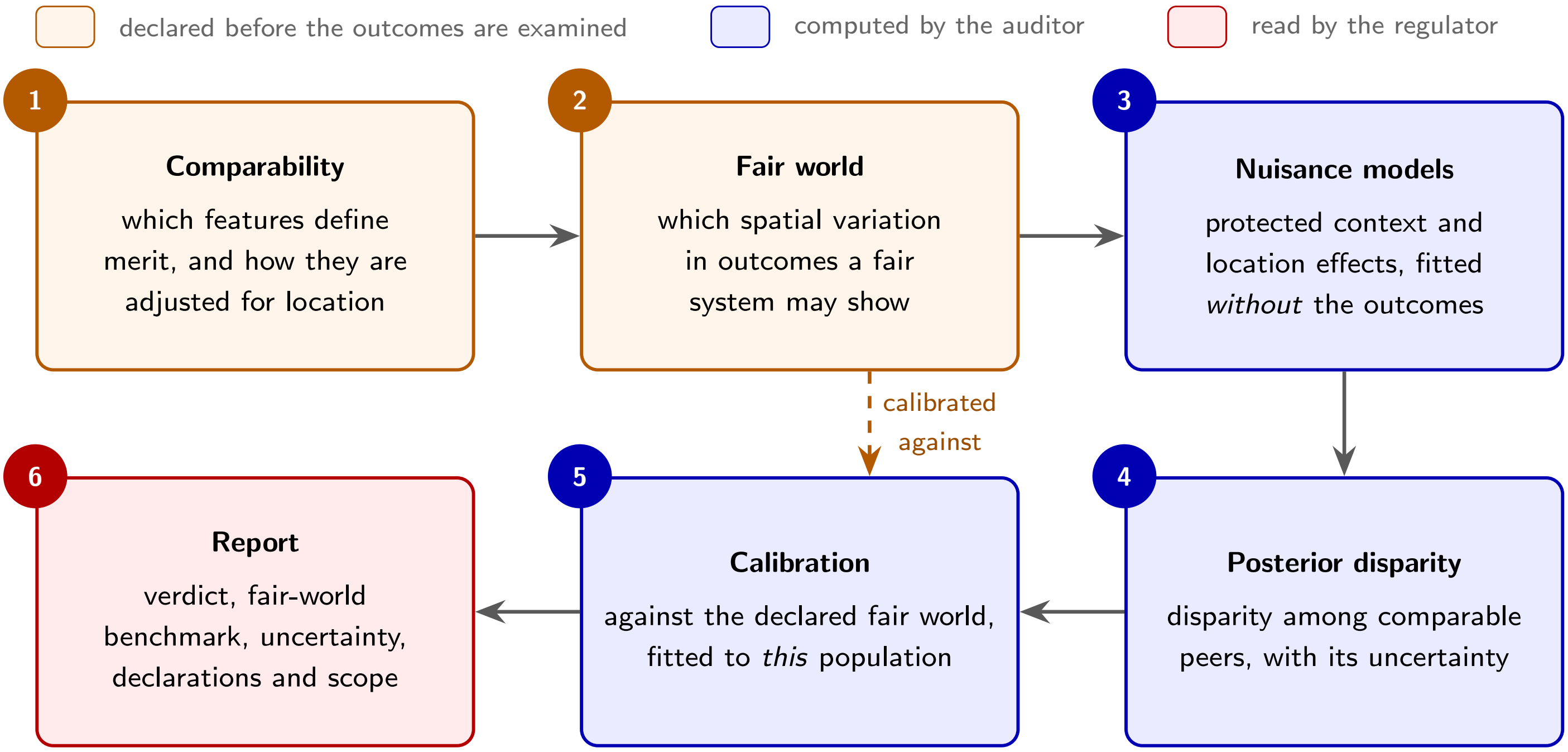
\begin{figure*}[htbp]
		\centering
		\resizebox{\linewidth}{!}{%
			\begin{tikzpicture}[
				font=\sffamily\small,
				node distance=0.9cm and 0.9cm,
				Step/.style={draw=#1!70!black, fill=#1!8, rounded corners=4pt, thick,
					align=center, text width=3.4cm, minimum height=2.3cm, inner sep=5pt,
					execute at begin node=\hyphenpenalty10000\exhyphenpenalty10000},
				num/.style={circle, fill=#1!70!black, text=white, font=\sffamily\bfseries\footnotesize,
					inner sep=1.4pt, minimum size=0.55cm},
				note/.style={font=\sffamily\footnotesize, text=black!60, align=center},
				key/.style={draw=#1!70!black, fill=#1!8, rounded corners=2pt, minimum width=0.5cm, minimum height=0.35cm},
				arr/.style={-{Stealth[length=2.6mm,width=2mm]}, line width=0.9pt, draw=black!65}
				]
				\node[Step=orange] (s1) {\textbf{Comparability}\\[3pt] which features define merit, and how they are adjusted for location};
				\node[Step=orange, right=of s1] (s2) {\textbf{Fair world}\\[3pt] which spatial variation in outcomes a fair system may show};
				\node[Step=blue, right=of s2] (s3) {\textbf{Nuisance models}\\[3pt] protected context and location effects, fitted \emph{without} the outcomes};
				\node[Step=blue, below=of s3] (s4) {\textbf{Posterior disparity}\\[3pt] disparity among comparable peers, with its uncertainty};
				\node[Step=blue, below=of s2] (s5) {\textbf{Calibration}\\[3pt] against the declared fair world, fitted to \emph{this} population};
				\node[Step=red, below=of s1] (s6) {\textbf{Report}\\[3pt] verdict, fair-world benchmark, uncertainty, declarations and scope};
				\foreach \n/\c/\k in {s1/orange/1, s2/orange/2, s3/blue/3, s4/blue/4, s5/blue/5, s6/red/6}
				\node[num=\c] at (\n.north west) {\k};
				\draw[arr] (s1) -- (s2);
				\draw[arr] (s2) -- (s3);
				\draw[arr] (s3) -- (s4);
				\draw[arr] (s4) -- (s5);
				\draw[arr] (s5) -- (s6);
				\draw[arr, dashed, draw=orange!70!black] ([xshift=6mm]s2.south) -- ([xshift=6mm]s5.north)
				node[midway, right, note, text=orange!60!black] {calibrated\\against};
				\node[key=orange, above=0.45cm of s1.north west, anchor=south west] (k1) {};
				\node[note, right=1mm of k1, anchor=west] (t1) {declared before the outcomes are examined};
				\node[key=blue, right=6mm of t1] (k2) {};
				\node[note, right=1mm of k2, anchor=west] (t2) {computed by the auditor};
				\node[key=red, right=6mm of t2] (k3) {};
				\node[note, right=1mm of k3, anchor=west] {read by the regulator};
			\end{tikzpicture}%
		}
		\caption{The six-step protocol, read from Step~1 to Step~6. Orange steps are normative declarations, fixed before the outcome data are examined; blue steps are statistical computations; the red step is what the regulator receives. The fair world declared in Step~2 is the model against which Step~5 calibrates, so a change to the declaration changes the verdict, and the declaration must therefore precede the data.}
		\label{fig:protocol}
	\end{figure*}

	An audit is a comparison between the system as observed and a fair world declared in advance. The protocol makes that comparison explicit. It has six steps, \fix{shown in Figure~\ref{fig:protocol} and Table~\ref{tab:protocol}}. Steps~1 and~2 are declarations: they encode normative judgements and must be fixed before the outcome data are examined, by the party with authority to make them. Steps~3 to~5 are computations performed by the auditor. Step~6 is the report. The steps are cumulative: each guards against a specific way in which an audit can mislead, and Section~\ref{sec:evidence} shows the size of each failure when its step is omitted.
	
	\begin{table*}[htbp]
		\centering
		\caption{The protocol. ``Failure prevented'' is the way in which an audit misleads when the step is omitted; ``Evidence'' points to the part of the simulation study that measures it.}
		\label{tab:protocol}
		\small
		\begin{tabularx}{\linewidth}{>{\raggedright\arraybackslash}p{2.2cm}>{\raggedright\arraybackslash}X>{\raggedright\arraybackslash}p{2.6cm}>{\raggedright\arraybackslash}p{3.6cm}>{\raggedright\arraybackslash}p{2.1cm}}
			\toprule
			Step & What is fixed & Who decides & Failure prevented & Evidence\\
			\midrule
			1. Comparability & The merit features that define who is comparable to whom, with a rationale for each; the adjustment of those features for location & Regulator with domain experts and affected parties & Comparing people who differ in legitimate merit, or treating location-driven advantage as merit & Design choice; Section~\ref{sec:bfair}\\
			2. Fair world & The class of outcome mechanisms declared fair, including how much spatial variation in outcomes is legitimate (the lattice side length $r$) & Regulator, before the data are examined & Legitimate clustering read as discrimination; or genuine disparity absorbed as ``legitimate'' & Section~\ref{sec:ev-fairworld}\\
			3. Nuisance models & Spatial models for the protected context and for the location effects in the features, fitted without the outcomes & Auditor & Outcome leakage into the statistic-defining nuisance quantities, which would invalidate the conditional exchangeability argument & \ig{Lemma~\ref{lem:cond}}\\
			4. Posterior disparity & The disparity among comparable peers as a posterior distribution, and the decision statistic & Auditor & A point score whose uncertainty is invisible & Section~\ref{sec:bfair}\\
			5. Calibration & The verdict, obtained by comparing the observed decision statistic with its distribution under the declared fair world in this population & Auditor & A threshold borrowed from another population & Section~\ref{sec:ev-threshold}\\
			6. Report & Decision, fair-world benchmark and its spread, posterior summaries, the declarations of Steps 1--2, and the scope of the conclusion & Auditor, to regulator & A verdict that cannot be checked or compared & Section~\ref{sec:worked}\\
			\bottomrule
		\end{tabularx}
	\end{table*}
	
	\paragraph{Step 1: Declare who is comparable}
	The audit compares individuals who are alike in the features that legitimately bear on the decision. Which features those are is not a statistical question. A feature may be predictive while encoding structural disadvantage \ig{\citep{barocas2016big}}, so each feature should be justified on substantive grounds and open to challenge by domain experts and affected parties\ig{ \citep{binns2018algorithmic}}. The protocol also takes a definitive stance on location-associated merit, removing the portion of a recorded feature explained by location before individuals are compared. This ensures that geography itself does not unfairly advantage or disadvantage individuals. A regulator who disagrees with this position must state so at this stage, as altering this premise fundamentally changes who is deemed comparable to whom.
	Crucially, the legitimacy of factoring in location depends heavily on the context. For instance, incorporating location is entirely legitimate when calculating flood insurance premiums, which logically cannot be identical for a desert property and an area prone to heavy rainfall. In this scenario, location reflects genuine physical risk. Conversely, using location becomes illegitimate when an individual's credit score is penalised simply because they reside in a low-income neighbourhood. In the latter case, geography merely acts as a proxy for systemic economic disadvantage.
	
	\paragraph{Step 2: Declare the fair world}
	A finding of unfairness indicates that the observed empirical distribution is statistically improbable under the null hypothesis of a fair system. Consequently the fair system must first be specified. In spatial data this specification contains a quantity that the outcome data cannot supply: how much spatial variation in outcomes a fair system may show. Two extremes bound the choice:
	\begin{itemize}
		\item \textbf{No spatial variation allowed.} A fair world in which outcomes depend on merit alone treats every regional difference in outcomes as potential unfairness, including differences with legitimate causes such as local labour markets or local risks.
		\item \textbf{Any spatial variation allowed.} A fair world that admits any spatial pattern absorbs the very disparity the audit is meant to detect, because the geography of the protected group is itself a spatial pattern.
	\end{itemize}
	The protocol places the fair world between these extremes. Legitimate spatial variation is represented by a smooth surface built from Gaussian bumps centred on an $r\times r$ lattice of knots (Section~\ref{sec:bfair}, Equation~\eqref{eq:fairworld}). The lattice side length $r$ is the declared parameter: on a unit-square domain the knots are $1/(r-1)$ apart, so $r=2$ admits only domain-wide gradients, while larger $r$ admits progressively finer regional variation. The choice of $r$ governs the error rates of the audit:
	\begin{itemize}
		\item \textbf{A coarser surface} (smaller $r$) leaves legitimate clustering outside the fair world. Benign regional differences are then read as unfairness, producing false findings.
		\item \textbf{A finer surface} (larger $r$) absorbs genuine disparity that varies on the same spatial scale as the protected group's geography, costing statistical sensitivity\ig{, that is, the probability that the audit flags a system as unfair when it is truly unfair}.
	\end{itemize}
	The trade-off cannot be removed statistically. In any one city, part of the legitimate spatial variation co-varies with the protected group's geography by chance, and that part is indistinguishable from aligned unfairness \citep{hodges2010adding,paciorek2010importance}; moreover, legitimate clustering and spatially structured disadvantage that does not track the protected context have the same statistical form, so whether a regional pattern is legitimate is a normative judgement, not an empirical one. More data sharpens the audit's view of the declared fair world, and exposes a declaration that is too coarse (Section~\ref{sec:ev-sweep}), but it cannot decide which variation ought to count as legitimate. The lattice side length is therefore an explicit operating point. A regulator should fix it, stated as the spatial scale of variation treated as legitimate, before the outcome data are examined, because a choice made afterwards could be tuned to produce any desired verdict.
	\paragraph{Step 3: Fit the nuisance models without the outcomes}
	The protected context (the probability that a resident of each location belongs to the protected group) and the location effects in the merit features are estimated from the protected attribute, the features and the locations, without using the outcomes under audit. These quantities define the residualised features, the peer graph and, ultimately, the statistic applied to each outcome vector. Keeping their estimation outcome-free allows them to be held fixed while only the outcome vector is regenerated in Step~5. This is the condition needed for the exchangeability argument of Lemma~\ref{lem:cond} when the fair-null probabilities are known. It is important to distinguish these statistic-defining nuisance models from the fair-null outcome model used in Step~5: the latter is fitted to the observed outcomes and therefore produces a plug-in calibration whose finite-sample level is approximate rather than exact.
	\paragraph{Step 4: \fix{Compute the posterior disparity}}
	\BFair{} yields a posterior distribution of disparity among comparable peers, not a single score. This allows the uncertainty coming from estimating the protected context and the peer relations to remain visible in the audit output. The decision is based on a lower posterior quantile, so that a system is flagged only when its disparity is large relative to its own uncertainty.

	\paragraph{Step 5: \fix{Calibrate against the audited population}}
	The verdict compares the observed decision statistic with its distribution when outcomes are regenerated from the declared fair world, fitted to the population being audited, with the nuisance models of Step~3 held fixed. No threshold is carried over from elsewhere: what is standardised is the procedure, not the number. The calibration requires no refitting of the spatial models.
	
	\paragraph{Step 6: \fix{Report the verdict and its basis}}
	The report states the verdict together with what is needed to read it: the declarations of Steps~1 and~2, the observed disparity and its posterior uncertainty, the fair-world benchmark for this population and its spread, the calibrated $p$-value, and the scope of the conclusion. A rejection is evidence of protected-context-aligned disparity relative to the declared fair world. It is not proof of a causal mechanism of discrimination, and it says nothing about spatial disadvantage that does not track the protected group's geography, which lies outside the estimand and should be examined separately.
	
	\fix{The protocol does not depend on the disparity measure used in Step~4. Section~\ref{sec:bfair} specifies the measure we use, \BFair{}, and the two forms that the calibration of Step~5 can take.}
	\section{\fix{The audit statistic and its calibration}}
	\label{sec:bfair}\label{sec:B-Fair}\label{sec:reference-calibration}\label{sec:conditional}
	\fix{This section specifies what Steps~3 to~5 compute: the disparity measure, the decision statistic derived from it, the benchmark a fair system produces, and the two ways in which a verdict can be calibrated.}
	
	\subsection{\fix{The disparity measure}}
	
	The full construction of \BFair{} \fix{is} given in \citet{jafta2026bfair} and
	Appendix~\ref{app:bfair}; this section states the quantities required
	to implement the audit. The observed data are
	\[
	D_n=\{(f_i,O_i,s_i,A_i)\}_{i=1}^n,
	\]
	where, for individual \(i\), \(f_i=(f_{i1},\ldots,f_{iq})^\top
	\in\mathbb R^q\) is a vector of \(q\) recorded merit features,
	\(O_i\in\{0,1\}\) is the audited outcome, with \(O_i=1\) denoting an
	adverse outcome, \(s_i\in\mathcal S\subset\mathbb R^2\) is the
	geographic location\ig{, with \(\mathcal S\) the study region covered by the audit}, and \(A_i\in\{0,1\}\) is the protected attribute,
	with \(A_i=1\) denoting membership of the protected group.
	
	Two sets of latent Gaussian models are fitted \ig{by the integrated nested Laplace approximation (INLA) \citep{rue2009inla,van2023new}, with each spatial field represented through the stochastic partial differential equation (SPDE) approach \citep{lindgren2011explicit}}. First, each recorded merit feature is decomposed into a global level, a location-associated
	component and an individual component. For \ig{each} feature \(k=1,\ldots,q\),
	the model is
	\[
	f_{ik}=\mu_k+u_k(s_i)+\varepsilon_{ik},
	\qquad
	\varepsilon_{ik}\sim\mathcal N(0,\sigma_k^2),
	\]
	where \(\mu_k\) is an intercept \ig{(a constant)} and \ig{\(u_k(\cdot)\)} is a zero-mean spatial Gaussian random field\ig{, a function of location evaluated at \(s_i\)}. For posterior draw \(m=1,\ldots,M\), the residualised feature is
	\[
	\widetilde f_{ik}^{(m)}
	=f_{ik}-\mu_k^{(m)}-u_k^{(m)}(s_i),
	\]
	and
	\[
	\widetilde f_i^{(m)}
	=
	\bigl(
	\widetilde f_{i1}^{(m)},\ldots,
	\widetilde f_{iq}^{(m)}
	\bigr)^\top
	\]
	is the corresponding residualised merit vector. Second, the protected attribute is modelled through the protected-context surface
	\[
	\pi(s)
	=
	\Pr(A=1\mid s),
	\qquad
	\operatorname{logit}\{\pi(s)\}
	=
	\beta_0+u_A(s),
	\]
	where \(\beta_0\) is an intercept and \ig{\(u_A(\cdot)\)} is a zero-mean \ig{spatial} Gaussian random field with Matérn covariance \ig{\citep{matern1986spatial,stein1999interpolation}}. Its practical range and marginal standard deviation receive penalised-complexity priors \citep{simpson2017penalising,fuglstad2019constructing}. \ig{The fitted model gives a joint posterior distribution for \(\beta_0\) and the field \(u_A(\cdot)\). Joint posterior draw \(m=1,\ldots,M\), written \(\beta_0^{(m)}\) and \(u_A^{(m)}(\cdot)\), defines one draw of the whole surface,
	\[
	\pi^{(m)}(s)=\operatorname{logit}^{-1}\bigl\{\beta_0^{(m)}+u_A^{(m)}(s)\bigr\},
	\]
	and evaluating this surface at the observed locations} gives the protected-context probabilities
	\[
	\pi_i^{(m)}
	:=
	\pi^{(m)}(s_i),
	\qquad i=1,\ldots,n.
	\]
	Thus, \(\pi_i^{(m)}\) is the estimated probability, in posterior draw
	\(m\), that a resident of the location occupied by individual \(i\)
	belongs to the protected group. Neither the feature models nor the
	protected-context model uses the audited outcomes \(O\), as required
	by Step~3.
	
	For each posterior draw \(m\), similarity in residualised merit is
	measured by the Mahalanobis distance \citep{mahalanobis1930tests}
	\begin{equation}
		d_{ij}^{(m)}
		=
		\left[
		\left\{
		\widetilde f_i^{(m)}-\widetilde f_j^{(m)}
		\right\}^{\!\top}
		\left(\widehat\Sigma^{(m)}\right)^{-1}
		\left\{
		\widetilde f_i^{(m)}-\widetilde f_j^{(m)}
		\right\}
		\right]^{1/2},
		\label{eq:mahalanobis}
	\end{equation}
	where \(\widehat\Sigma^{(m)}\) is the empirical covariance matrix of
	\(\widetilde f_1^{(m)},\ldots,\widetilde f_n^{(m)}\), with a small
	ridge term added before inversion when required for numerical
	stability.
	
	A directed \(k\)-nearest-neighbour graph is first constructed using the distances \(d_{ij}^{(m)}\). It is then symmetrised by union: an unordered pair \(\{i,j\}\) is retained if \(j\) is among the \(k\) nearest neighbours of \(i\), or \(i\) is among the \(k\) nearest neighbours of \(j\). Let
	\[
	E^{(m)}
	=
	\bigl\{
	(i,j):i<j\ \text{and the pair }\{i,j\}
	\text{ is retained}
	\bigr\}
	\]
	denote the resulting set of peer edges. Each retained pair receives
	the Gaussian-kernel weight
	\begin{equation}
		w_{ij}^{(m)}
		=
		\exp\left\{
		-\frac{\bigl(d_{ij}^{(m)}\bigr)^2}
		{2\bigl(h^{(m)}\bigr)^2}
		\right\},
		\qquad (i,j)\in E^{(m)},
		\label{eq:peer-weight}
	\end{equation}
	where \(h^{(m)}>0\) is the kernel bandwidth. In the implementation used
	here, \(h^{(m)}\) is the empirical median of the Mahalanobis distances
	over the retained edges in \(E^{(m)}\). Pairs outside \(E^{(m)}\) are
	not included in the disparity functional.
	Directionality is encoded by
	\begin{equation}
		\Delta_{ij}^{(m)}
		=
		\mathbf 1\{\pi_i^{(m)}>\pi_j^{(m)}\}
		O_i(1-O_j)
		+
		\mathbf 1\{\pi_j^{(m)}>\pi_i^{(m)}\}
		O_j(1-O_i),
		\label{eq:direction-indicator}
	\end{equation}
	where \(\mathbf 1\{\cdot\}\) is the indicator function. Consequently,
	\(\Delta_{ij}^{(m)}=1\) precisely when the pair has discordant outcomes
	and the individual living in the higher protected-context location
	receives the adverse outcome. It equals zero when the outcomes agree,
	when the adverse outcome occurs in the lower protected-context
	location, or when the two protected-context values are equal.
	The directional disparity for posterior draw \(m\) is
	\begin{equation}
		\Tdir^{(m)}
		=
		\frac{
			\displaystyle
			\sum_{(i,j)\in E^{(m)}}
			w_{ij}^{(m)}
			\left|\pi_i^{(m)}-\pi_j^{(m)}\right|
			\Delta_{ij}^{(m)}
		}{
			\displaystyle
			\sum_{(i,j)\in E^{(m)}}w_{ij}^{(m)}
		}.
		\label{eq:tdir}
	\end{equation}
	The denominator is assumed to be positive, as it is whenever the peer
	graph contains at least one retained edge. Because every term in the
	numerator is non-negative and
	\(\left|\pi_i^{(m)}-\pi_j^{(m)}\right|\leq 1\), the statistic satisfies
	\(0\leq\Tdir^{(m)}\leq1\).
	
	A retained pair therefore contributes to the disparity when two
	merit-comparable individuals live in different protected contexts and
	the individual in the higher protected context receives the adverse
	outcome. Its contribution increases with both merit similarity,
	through \(w_{ij}^{(m)}\), and the difference in protected context,
	through
	\(\left|\pi_i^{(m)}-\pi_j^{(m)}\right|\). The statistic is consequently
	large when strongly comparable pairs with substantially different
	protected contexts are discordant in the unfavourable direction. It is
	small when peers live in similar protected contexts, when their
	outcomes agree, or when discordance runs in the opposite direction.
	
	Repeating the calculation for \(m=1,\ldots,M\) produces the empirical
	posterior distribution
	\[
	\bigl\{
	\Tdir^{(1)},\ldots,\Tdir^{(M)}
	\bigr\},
	\]
	which propagates uncertainty in the residualised merit features, the
	peer relations and the protected-context surface through the
	directional disparity functional. This is the posterior-disparity
	calculation in Step~4.
	
	For a prespecified posterior tail level
	\(\alpha\in(0,0.5)\), the scalar decision statistic is the lower
	posterior quantile
	\begin{equation}
		\Tdec
		=
		Q_\alpha\!\left(
		\Tdir\mid D_n
		\right),
		\label{eq:tdec}
	\end{equation}
	where
	\[
	Q_\alpha(X)
	=
	\inf\{t:\Pr(X\leq t)\geq\alpha\}.
	\]
	In computation, \(\Tdec\) is the empirical \(\alpha\)-quantile of
	\(\Tdir^{(1)},\ldots,\Tdir^{(M)}\). For example, when
	\(\alpha=0.05\) and \(M=1000\), the implementation uses the relevant
	lower-tail order statistic specified by the adopted empirical-quantile
	convention. The posterior tail level \(\alpha\) controls how strongly
	the posterior distribution must support a large disparity; it is
	distinct from the audit's false-rejection level, which is determined
	by the fair-null calibration in Step~5.
	
\begin{figure*}[htbp]
		\centering
		\resizebox{\linewidth}{!}{%
			\begin{tikzpicture}[
				font=\sffamily\small,
				node distance=0.7cm and 0.75cm,
				stage/.style={draw=#1!70!black, fill=#1!8, rounded corners=4pt, thick,
					align=center, text width=3.1cm, minimum height=2.15cm, inner sep=5pt,
					execute at begin node=\hyphenpenalty10000\exhyphenpenalty10000},
				stage/.default=gray,
				num/.style={circle, fill=#1!70!black, text=white, font=\sffamily\bfseries\footnotesize,
					inner sep=1.4pt, minimum size=0.5cm},
				note/.style={font=\sffamily\footnotesize, text=black!55},
				arr/.style={-{Stealth[length=2.6mm,width=2mm]}, line width=0.9pt, draw=black!65}
				]
				\node[stage=blue] (merit) {\textbf{Adjust merit for location}\\[3pt]
					remove the part of each feature that is due to \emph{where} a person lives};
				\node[stage=orange, below=0.6cm of merit] (ctx) {\textbf{Map the protected context}\\[3pt]
					a smooth surface of group composition across the region};
				\path (merit.west) -- (ctx.west) coordinate[midway] (mid);
				\node[stage=gray, left=1.1cm of mid, anchor=east] (data) {\textbf{Audit data}\\[3pt]
					for each person: location, merit features, decision, protected group};
				\node[stage=teal, right=of merit] (peers) {\textbf{Find comparable peers}\\[3pt]
					link people with similar adjusted merit; closer peers count more};
				\node[stage=teal, right=of peers] (score) {\textbf{Score unfavourable pairs}\\[3pt]
					peers in different contexts where the one in the more protected area loses out};
				\node[stage=violet, anchor=west] (post) at ([xshift=1.0cm]mid -| score.east) {\textbf{Disparity with uncertainty}\\[3pt]
					one score per posterior draw gives a distribution, not a single number};
				\node[stage=red, right=of post] (verdict) {\textbf{Verdict}\\[3pt]
					flag only if the disparity clearly exceeds what a \emph{fair} system here would produce};
				\node[stage=gray, fill=white, dashed, below=0.7cm of verdict, minimum height=1.5cm] (ref)
				{\textbf{Fair-world reference}\\[2pt]{\footnotesize same people and places,  outcomes driven by declared legitimate factors: residualised merit plus the declared spatial component}};
				\foreach \n/\c/\k in {data/gray/1, merit/blue/2a, ctx/orange/2b, peers/teal/3, score/teal/4, post/violet/5, verdict/red/6}
				\node[num=\c] at (\n.north west) {\k};
				\draw[arr] (data.east) -- ++(0.55,0) |- (merit.west);
				\draw[arr] (data.east) -- ++(0.55,0) |- (ctx.west);
				\draw[arr] (merit) -- (peers);
				\draw[arr] (peers) -- (score);
				\draw[arr] (ctx.east) -| (score.south) node[pos=0.25, note, below] {differences in protected context};
				\draw[arr] (score.east) -- ++(0.35,0) |- (post.west);
				\draw[arr] (post) -- (verdict);
				\draw[arr, dashed] (ref) -- (verdict) node[midway, note, right] {threshold};
				\begin{scope}[on background layer]
					\node[draw=black!40, dashed, rounded corners=8pt, inner sep=8pt, fill=black!2,
					fit=(merit)(ctx)(peers)(score)] (loop) {};
				\end{scope}
				\node[note, anchor=south west] at ([xshift=4pt]loop.north west) {\textbf{repeated for every posterior draw}};
			\end{tikzpicture}%
		}
		\caption{How \BFair{} reaches a verdict. (1)~The audit starts from individual records. (2)~Two spatial models separate what is due to location: (2a)~merit features are adjusted for where people live, and (2b)~the protected composition of each area is mapped as a smooth surface. (3)~Individuals with similar adjusted merit are linked as peers, closer peers receiving more weight. (4)~A peer pair counts against the system when the two live in different protected contexts and the person in the more protected area receives the adverse outcome. Steps 2--4 are repeated for every posterior draw($m$), so (5)~the disparity is a posterior distribution rather than a single number. (6)~A system is flagged only when the disparity exceeds, with high posterior probability, a threshold calibrated against a fair-world reference for the same population (Section~\ref{sec:reference-calibration}). In Part~I the spatial quantities are supplied at their true values and the loop is run once (Section~\ref{sec:part1-decision-mechanisms}).}
		\label{fig:bfair-logic}
	\end{figure*}
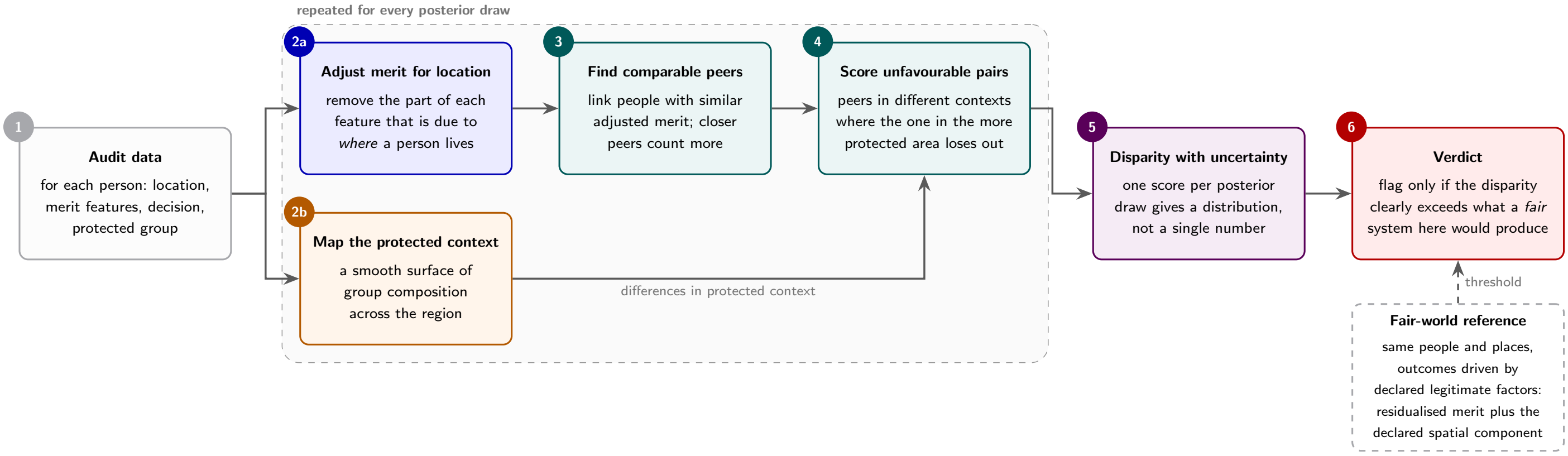

	Figure~\ref{fig:bfair-logic} summarises the construction of the
	posterior disparity and its reduction to the decision statistic
	\(\Tdec\). Step~5 then compares the observed value of \(\Tdec\) with
	the distribution obtained by regenerating the outcomes under the
	declared fair-null model while holding the statistic-defining
	quantities fixed.
	
	\subsection{\fix{The decision statistic}}
	\chg{For a posterior tail level $\alpha$, the decision statistic is the lower $100\alpha\%$ posterior quantile, $\Tdec=Q^+_{\alpha}\{\Tdir\mid\mathcal D\}$; with $\alpha=0.05$ and $M=1000$ draws it is the 51st smallest draw. For any threshold $\tau$, $\Tdec>\tau$ holds exactly when $\widehat\Pr(\Tdir>\tau\mid\mathcal D)\ge1-\alpha$, or equivalently $\widehat\Pr(\Tdir\le\tau\mid\mathcal D)\le\alpha$, so a system is flagged only when, under the fitted model, its disparity exceeds the benchmark with posterior probability at least $1-\alpha$ . The tail level $\alpha$ is distinct from the audit's false-rejection rate, which is controlled by the calibration of Step~5.}
	
	\subsection{\fix{The fair-null pedestal: why zero is the wrong benchmark}}\label{sec:sim-implications}
	Even in a fair system, chance produces comparable pairs in which the person in the more protected context receives the adverse outcome, so $\Tdir$ is positive under fairness. When outcomes are independent with fair probabilities $p_{0,i}$, its expected value is
	\begin{equation}
		\begin{aligned}
			\theta_0(\mathcal D) &= \frac{\sum_{i<j} w_{ij}\,|\pi_i-\pi_j|\,c_{ij}}{\sum_{i<j}w_{ij}}, \\
			\text{where } c_{ij} &= p_{0,i}(1-p_{0,j})\,\mathbf{1}(\pi_i>\pi_j) + p_{0,j}(1-p_{0,i})\,\mathbf{1}(\pi_j>\pi_i).
		\end{aligned}
		\label{eq:pedestal}
	\end{equation}
	which we call the \emph{fair-null pedestal}. The term $c_{ij}$ isolates the core stochastic mechanism behind this phenomenon, representing the joint probability that a pair $(i, j)$ yields a naturally inverted outcome purely by chance. Since the individual outcomes are independent under fairness, this joint probability is the product of their independent probabilities. The indicator functions $\mathbf{1}(\cdot)$ ensure that only the contextually relevant term is active: if individual $i$ is more protected ($\pi_i > \pi_j$), $c_{ij}$ simplifies to the probability that $i$ suffers the adverse outcome ($p_{0,i}$) while $j$ avoids it ($1-p_{0,j}$), and vice versa. 
	
	As a result, the fair-null pedestal depends on who lives where, who is compared with whom, and each person's fair outcome probability, causing it to shift dynamically from one audited population to the next. This formalises why zero is the wrong benchmark, and why a fixed threshold is an incorrect standard for auditing fairness.
	
	\subsection{\fix{The declared fair world and the calibrated verdict}}
	\ig{Recall that} $\widetilde f_i\in\mathbb R^q$ \ig{denotes} the vector of
	location-residualised merit features for individual $i$ \ig{(Section~\ref{sec:bfair})}, and
	$s_i\in\mathcal S\subset\mathbb R^2$ \ig{denotes} that individual's observed
	location. In Step~2, the regulator declares the following restricted
	spatial logistic model to represent a fair outcome mechanism:
	\begin{equation}
		\operatorname{logit} p^{(0,r)}_i
		=
		\gamma_0
		+
		\widetilde f_i^{\top}\boldsymbol\gamma
		+
		\widetilde{\mathbf B}_r(s_i)^{\top}\boldsymbol\zeta_r,
		\qquad i=1,\ldots,n,
		\label{eq:fairworld}
	\end{equation}
	where $p^{(0,r)}_i=\Pr(O_i=1\mid\widetilde f_i,s_i;H_0^{(r)})$
	is the fair-world probability of the adverse outcome $O_i=1$,
	$\gamma_0\in\mathbb R$ is an intercept,
	$\boldsymbol\gamma\in\mathbb R^q$ contains the merit coefficients,
	and $\boldsymbol\zeta_r\in\mathbb R^{K_r}$ contains the spatial-basis
	coefficients. The vector
	$\widetilde{\mathbf B}_r(s_i)\in\mathbb R^{K_r}$ is obtained from
	Gaussian radial-basis functions centred on a regular $r\times r$
	lattice of knots spanning the observed coordinate ranges; the basis
	columns are centred and reduced to full numerical rank, so that
	$K_r\leq r^2$. The protected-context surface $\pi(s_i)$, the realised
	protected attribute $A_i$, and any direct proxies for either are
	excluded from Equation~\eqref{eq:fairworld}. Increasing the declared
	lattice side length $r$ reduces the knot spacing and allows the
	fair-world model to represent finer residual spatial variation in
	outcomes.
	
	In Step~5, the auditor fits Equation~\eqref{eq:fairworld} to the
	observed outcomes, obtaining fitted fair-world probabilities
	$\widehat p^{(0,r)}_1,\ldots,\widehat p^{(0,r)}_n$. For
	$b=1,\ldots,B$, the auditor independently generates a complete outcome
	vector
	\[
	O_b^\star=(O_{b1}^\star,\ldots,O_{bn}^\star),
	\qquad
	O_{bi}^\star\mid\widehat p^{(0,r)}_i
	\sim
	\operatorname{Bernoulli}\!\left(\widehat p^{(0,r)}_i\right),
	\]
	with independence across individuals $i$ and replicates $b$. For each
	$O_b^\star$, the auditor recomputes the same decision statistic
	\[
	T_{\mathrm{dec}}(O_b^\star)
	=
	\fix{Q^{+}_{\alpha}}\!\left\{
	T_{\mathrm{dir}}(O_b^\star)\mid\mathcal D_b^\star
	\right\},
	\]
	where $\fix{Q^{+}_{\alpha}}$ denotes the lower \fix{(right-continuous)} posterior $\alpha$-quantile,
	$\alpha=0.05$ in the reported analysis, and
	$\mathcal D_b^\star$ differs from the observed data only through the
	replacement of $O$ by $O_b^\star$. The posterior draws of the
	protected-context surface and residualised features, together with the
	associated peer graphs and weights, are held fixed at those obtained
	without using the observed outcomes.
	
	Writing
	\[
	V_0=T_{\mathrm{dec}}(O)
	\qquad\text{and}\qquad
	V_b=T_{\mathrm{dec}}(O_b^\star),\quad b=1,\ldots,B,
	\]
	the upper-tail Monte Carlo $p$-value is
	\begin{equation}
		p_{\mathrm{cond}}
		=
		\frac{1+\sum_{b=1}^{B}\mathbf 1\{V_b\geq V_0\}}{B+1}.
		\label{eq:pcond}
	\end{equation}
	The declared fair world $H_0^{(r)}$ is rejected, and the system is
	flagged, when $p_{\mathrm{cond}}\leq\fix{\varepsilon}$, with audit significance
	level $\fix{\varepsilon}=0.05$. The posterior tail level $\alpha$, which defines
	$T_{\mathrm{dec}}$, and the audit significance level $\fix{\varepsilon}$, which
	defines the final verdict, therefore have distinct roles.
	
	For the realised audited population, the reported fair-world
	benchmark is
	\[
	\widehat\mu_{0,r}
	=
	\frac{1}{B}\sum_{b=1}^{B}V_b,
	\]
	the Monte Carlo estimate of the conditional mean of the decision
	statistic under the fitted declared fair world. Since $\widehat p^{(0,r)}_i$ is estimated from the same observed outcomes being tested, this operational procedure is a plug-in parametric bootstrap: $p_{\mathrm{cond}}$ is exactly conditionally valid only in the ideal case in which the fair-world probabilities are fixed independently of $O$, and otherwise has approximate finite-sample calibration.

	\begingroup\fixcolor
	\subsection{\fix{Transported and design-conditional calibration}}
	\ig{Throughout, a \emph{city} means one audited population: $n$ individuals in a study region, with their locations, protected attributes, recorded merit features and decision outcomes. A regulator audits one city at a time; in the simulation study of Section~\ref{sec:evidence} each simulated dataset is one city.} The decision statistic $\Tdec$ is the same in every audit; what differs is the reference against which it is judged. Two references are compared in this paper, and they answer different questions. Write $\mathcal D=(s_i,A_i,f_i)_{i=1}^n$ for the design of a city, that is, everything except the outcomes, and $\mathcal G$ for the design together with the posterior draws of the protected context, the residualised features and the peer graph, all of which are computed without the outcomes.
	
	\begin{definition}[Transported threshold]\label{def:transport}
		Let $P_0$ be a fair data-generating mechanism over cities, specifying the distribution of the design $\mathcal D$ and of the outcomes given $\mathcal D$. Simulate $\Bcal$ cities from $P_0$, fit the nuisance models and compute $\Tdec^{(c)}$ for each city $c=1,\dots,\Bcal$, and set
		\[
		\tau_{P_0}=\text{the empirical }(1-\varepsilon)\text{ quantile of }\Tdec^{(1)},\dots,\Tdec^{(\Bcal)}.
		\]
		The threshold is computed once. Any audited city is flagged when $\Tdec>\tau_{P_0}$.
	\end{definition}
	
	\begin{definition}[Design-conditional calibration]\label{def:cond}
		For the audited city, fit the declared fair world Equation~\eqref{eq:fairworld} to its outcomes, draw $O^\star_1,\dots,O^\star_B$ from the fitted probabilities, recompute $\Tdec(O^\star_b)$ with $\mathcal G$ held fixed, and compute $p_{\mathrm{cond}}$ by Equation~\eqref{eq:pcond}. The city is flagged when $p_{\mathrm{cond}}\le\varepsilon$. The reference is recomputed for every audit.
	\end{definition}
	
	The two definitions control different error rates. The transported threshold controls the false-rejection rate \emph{averaged over cities drawn from the calibration mechanism},
	\[
	\Pr_{P_0}\bigl(\Tdec>\tau_{P_0}\bigr)\approx\varepsilon,
	\]
	where the probability is over the design and the outcomes jointly. It says nothing about a city drawn from another fair mechanism $P_1$, and nothing about any particular city. The design-conditional calibration targets the false-rejection rate \emph{for the city being audited},
	\[
	\Pr\bigl(p_{\mathrm{cond}}\le\varepsilon\mid\mathcal G\bigr)\le\varepsilon,
	\]
	where the probability is over the outcomes alone, with the city's design held fixed. This bound is exact when the fitted probabilities equal the true fair probabilities (Lemma~\ref{lem:cond}) and approximate otherwise. Because the bound is conditional on the design, it does not depend on the population from which the city came, to the extent that the declared fair world is adequate for that city.
	
	The decomposition of the variance of the statistic across fair cities,
	\[
	\Var(\That)=\E\{\Var(\That\mid\mathcal D)\}+\Var\{\theta_0(\mathcal D)\},
	\]
	shows why the difference matters. A transported threshold absorbs both terms as they were under $P_0$; if the between-city term, the spread of the pedestal $\theta_0(\mathcal D)$, is larger in the audited population than in the calibration population, fair cities with a high pedestal cross the threshold. The design-conditional reference is centred on the audited city's own fair-world benchmark, so the between-city term does not enter. Table~\ref{tab:calib} summarises the contrast.
	
	\begin{table}[htbp]
		\centering\fixcolor
		\caption{The two calibrations compared. Rates are for the feasible plugin audit; ``B3'' is fair with benign clustering, ``B16'' is unfair with benign clustering (Section~\ref{sec:evidence})\ig{; \emph{benign clustering} is spatial clustering of outcomes that is unrelated to merit and to the protected group, and so is not unfair}.}
		\label{tab:calib}
		\footnotesize\setlength{\tabcolsep}{4pt}
		\begin{tabularx}{\linewidth}{>{\raggedright\arraybackslash}p{0.27\linewidth}>{\raggedright\arraybackslash}X>{\raggedright\arraybackslash}X}
			\toprule
			& Transported threshold & Design-conditional calibration\\
			\midrule
			Reference distribution & $\Tdec$ across fair cities from $P_0$ & $\Tdec$ across resampled outcomes in the audited city\\
			Computed & once, for all audits & for each audit\\
			Model refits & $\Bcal$ fits of the nuisance models, before any audit & none; $B\times M$ evaluations of $\Tdir$\\
			Requires & a simulator of fair cities & a declared fair world for this city (lattice side length $r$)\\
			Error rate controlled & averaged over cities from $P_0$ & for the audited city\\
			Main source of error & audited city not drawn from $P_0$ & lattice approximation; $\widehat p_0$ estimated from the tested outcomes\\
			False rejection, B3 & 9.2\% & 3.4\%\\
			Detection, B16 & 27.4\% & 22.4\%\\
			\bottomrule
		\end{tabularx}
	\end{table}
	\endgroup
	
	\fix{Section~\ref{sec:evidence} asks how much each of these choices matters, by auditing simulated cities whose fairness status is known.}
	
	\section{\fix{Evidence: what goes wrong when a step is omitted}}
	\label{sec:evidence}\label{sec:sim-results}
	\fix{We take the steps in the order of what they protect. Steps~2 and~5 protect validity, the rate at which fair systems are wrongly flagged (Sections~\ref{sec:ev-fairworld} and~\ref{sec:ev-threshold}); Step~3 governs sensitivity, the rate at which unfair systems are detected (Section~\ref{sec:ev-estimation}). Section~\ref{sec:ev-sweep} asks whether the conclusions hold beyond the configuration studied.}
	
	\subsection{\fix{Study design and what each audit knows}}
	The operating characteristics of an audit cannot be established from real data, because the true fairness status of a real decision system is unknown. We therefore build artificial cities, make the decision system fair in some and unfair in others, run the audit without telling it which is which, and count how often it is right. \ig{Each simulated dataset is one city in the sense of Section~\ref{sec:conditional}: $n$ individuals generated from mechanisms we control, so that its true fairness status is known.} Each mechanism that can produce spatially patterned outcomes enters through its own term and can be switched on or off (Table~\ref{tab:mechanisms}); Appendix~\ref{app:dgp} gives the full specification.
	
	\begin{table*}[tbp]
		\centering
		\caption{Mechanisms represented in the data-generating process.
			``Target'' indicates whether the mechanism constitutes
			protected-context-aligned unfairness, the estimand \BFair{} is designed to
			detect.}
		\label{tab:mechanisms}
		\small
		\begin{tabularx}{\linewidth}{>{\raggedright\arraybackslash}p{2.9cm}>{\raggedright\arraybackslash}p{3.1cm}X c}
			\toprule
			Mechanism & Generating term & Substantive content & Target\\
			\midrule
			Residential segregation & $a_\pi z_\pi(s)$ &
			The protected group is concentrated in some areas. & --\\
			Location-associated merit & $a_{fk}z_{fk}(s)+g_k(s)$ &
			Recorded merit features vary with location, for example because income or
			credit history differs between neighbourhoods. & No\\
			Benign spatial outcome dependence & $\aout\zout(s)$ &
			Outcomes cluster in space for reasons unrelated to merit and to the
			protected group. & No\\
			Aligned unfairness & $\dd\,\kalign\,\pic(s)$ &
			Merit-comparable individuals in higher-protected-context settings are more
			likely to receive the adverse outcome. & Yes\\
			Misaligned spatial unfairness & $\dd\,(1-\kalign)\,\uc(s)$ &
			A spatially structured disadvantage that does not track the protected
			context. & No\\
			\bottomrule
		\end{tabularx}
	\end{table*}
	
	Each city has $n=800$ residents placed uniformly on the unit square. The protected group is concentrated in some areas (a Matérn surface of practical range $0.25$). Recorded merit features carry location-associated advantage, which the audit must remove. In the stress scenario, outcomes also carry a benign spatial field of comparable range ($0.30$) that is independent of the protected group's geography, the setting in which a fair world can least easily separate legitimate clustering from unfairness. The injected unfairness has magnitude $\dd=1.25$ on the log-odds scale. Three mechanisms are compared, each differing from the last by one switch (Table~\ref{tab:design}), and each is audited twice: in an oracle analysis (Part~I), which is given the true protected context and residualised features and so isolates the reference and the threshold, and in a feasible Bayesian analysis (Part~II), which estimates them by SPDE--INLA as a real audit would. The feasible analysis of the unfair mechanism is run on the very datasets used in its oracle analysis. \ig{Each combination of mechanism and analysis, a cell of Table~\ref{tab:design}, is run on 500 simulated cities} (1500 and 1000 for the pooled oracle fair cells).

	\begin{table*}[htbp]
		\centering
		\caption{The three mechanisms. Within a row the oracle (P) and feasible (B) cells share the data-generating process; the B16 cell is run on the P4 datasets themselves.}
		\label{tab:design}
		\small
		\begin{tabular}{lcccc}
			\toprule
			Mechanism & $\dd$ & $\aout$ & Oracle analysis & Feasible analysis\\
			\midrule
			Fair, no benign clustering   & 0    & 0    & P1 & B1 (threshold calibration)\\
			Fair, benign clustering      & 0    & 0.80 & P3 & B3\\
			Unfair, benign clustering    & 1.25 & 0.80 & P4 & B16 (same data as P4)\\
			\bottomrule
		\end{tabular}
	\end{table*}
	\paragraph{What each audit knows} In every simulated city the true fair outcome probabilities are known to us, because we generated them: with $\dd=0$, $\operatorname{logit}p_{0,i}=\tilde f_i^{\top}\boldsymbol\beta_{\mathrm{merit}}+\aout\zout(s_i)$, where $\tilde f_i$ is the true location-free merit and $\zout$ the benign field. The audits differ in how much of this they are given and in how they build the fair-world reference against which the observed $\Tdec$ is judged. Table~\ref{tab:knows} sets this out. Two sources of error are kept apart throughout. \emph{Estimation} error arises when the protected context and residualised merit must be inferred from the data. \emph{Specification} error arises when the declared fair world cannot represent the true fair mechanism, whatever the data. The oracle analysis has no estimation error, so any failure it shows is a failure of specification or of the threshold.

	\ig{Figure~\ref{fig:city} shows one simulated city under three of these mechanisms, for the same individuals and the same random numbers, so that every difference between its panels is due to the mechanism alone. The benign field in panel~(b) is fair by construction, yet it is spatially smooth on a scale comparable to the protected context in panel~(a) (ranges $0.30$ and $0.25$), which is what makes it hard to separate from unfairness. Panel~(c) shows the consequence. Pure aligned unfairness raises the adverse probability steadily with $\pi$; this gradient is the signal \BFair{} is designed to detect. The benign field alone should show no gradient on average over cities, but in this city it happens to be lower where $\pi$ is high. When both are present this chance alignment cancels the unfairness and the gradient almost disappears. In other cities the alignment has the opposite sign and makes a fair city look unfair. This city-to-city variation in chance alignment is what moves the fair-null pedestal between populations (Section~\ref{sec:sim-implications}). Panels~(d)--(f) show that the unfairness term, the difference between (f) and (e), is small relative to the benign field, so P3 and P4 look almost identical by eye; this is why the P3/P4 pair is a demanding test of both validity and sensitivity.}
	
	\begin{table*}[htbp]
		
		\caption{How each audit builds its reference, what it is given, and where it departs from the truth. ``True'' means supplied at the simulated value. In every audit the outcomes are resampled with the protected context, residualised merit and peer graph held fixed, and $\Tdec$ is recomputed for each resampled outcome vector.}
		\label{tab:knows}
		\footnotesize\setlength{\tabcolsep}{4pt}
		\begin{tabularx}{\linewidth}{>{\raggedright\arraybackslash}p{0.15\linewidth}>{\raggedright\arraybackslash}p{0.13\linewidth}>{\raggedright\arraybackslash}X>{\raggedright\arraybackslash}p{0.33\linewidth}}
			\toprule
			Audit & Protected context $\pi$ and merit $\tilde f$ & Reference against which $\Tdec$ is judged & Exact, or where it is misspecified\\
			\midrule
			\multicolumn{4}{l}{\emph{Oracle analysis (Part~I): $\Tdec=\That$, one value per city; 499 resamples per city}}\\
			True fair world & true & outcomes redrawn from the true $p_{0,i}$ & \textbf{exact}; infeasible, since no auditor knows $p_{0,i}$\\
			Merit-only fair world & true & logistic regression of $O$ on $\tilde f$, fitted to the city; outcomes redrawn from the fit & correctly specified without benign clustering; with it, \textbf{omits the benign field} $\aout\zout$, so resampled cities are too uniform\\
			Spatial fair world, $r$ & true & logistic regression of $O$ on $\tilde f$ and an $r\times r$ lattice surface Equation~\eqref{eq:fairworld}; outcomes redrawn from the fit & the lattice surface \textbf{approximates} the benign field but cannot reproduce a Matérn field exactly for any finite $r$; the residual misfit becomes detectable as $n$ grows, and when $\dd>0$ the surface also absorbs part of $\pi$\\
			\midrule
			\multicolumn{4}{l}{\emph{Feasible analysis (Part~II): $\Tdec$ from 1000 posterior draws}}\\
			Benchmark, transported & true & one threshold $\tau$: the 95th percentile of $\Tdec$ over 2000 fair cities \emph{without} benign clustering & \textbf{wrong population}: the threshold encodes a fair world unlike the audited one\\
			Plugin or full, transported & estimated by SPDE--INLA: $\pi$ from $A$; location effects from the features & the same kind of threshold, computed for this variant & wrong population, plus estimation error: $\widehat\pi$ is shrunk towards its mean, and the stationary SPDE model \textbf{cannot represent} the deterministic trends $g_k$ in the features\\
			Benchmark, conditional & true & spatial fair world ($r=3$) fitted to this city; 499 resampled outcome vectors & lattice approximation only\\
			Plugin, conditional & estimated, as above; draws held fixed & spatial fair world ($r=3$) fitted on posterior-mean $\tilde f$; 499 resampled outcome vectors & lattice approximation, estimation error, and $\widehat p_0$ \textbf{estimated from the same outcomes} that are tested (plug-in)\\
			\bottomrule
		\end{tabularx}
	\end{table*}

	\begin{figure*}[tbp]
		\centering
		\figslot{figures/fig6_3_city.pdf}{8cm}
		\caption{\ig{One simulated city ($n=800$) under three outcome mechanisms, for the same individuals and the same random numbers.
			(a)~Protected context $\pi(s)$; white dots mark protected-group members.
			(b)~Benign outcome field $\zout(s)$, generated independently of $\pi$ and of merit.
			(c)~For each quintile of protected context (160 individuals each), the mean outcome probability under each mechanism minus its mean under P1; purple: aligned unfairness only ($\dd=1.25$, $\aout=0$); blue: P3; red: P4.
			(d)--(f)~The location-dependent part of the outcome log-odds, $\aout\zout(s)+\dd\,\pic(s)$, under P1, P3 and P4, with adverse outcomes as black dots.}}
		\label{fig:city}
	\end{figure*}
	
	\begin{table*}[htbp]
		\centering
		\caption{Rejection rates (\%) with 95\% intervals. Rows 1--2 are fair mechanisms, so every entry is a false-rejection rate; row 3 is a sensitivity. Part~I: 1500 (P1) and 1000 (P3) datasets, each pooled over design-equivalent cells on independent streams, and 500 (P4); Part~II: 498--500 datasets per cell, the B16 cell on the P4 datasets. Conditional: design-conditional calibration at the primary lattice $r=3$. In the Part~I lattice family R7 (500 datasets per arm), the restricted-spatial reference at $r=3$ has false-rejection rate 4.4\% and sensitivity 46.6\%.}
		\label{tab:main}
		\setlength{\tabcolsep}{3pt}
		\resizebox{\linewidth}{!}{%
			\begin{tabular}{lccc|ccc|cc}
				\toprule
				& \multicolumn{3}{c|}{Part I (Monte Carlo $p$-value)} &
				\multicolumn{3}{c|}{Part II, transported threshold} &
				\multicolumn{2}{c}{Part II, conditional}\\
				Mechanism & merit-only & spatial, $r=2$ & oracle & benchmark & plugin & full & benchmark & plugin\\
				\midrule
				Fair, $\aout=0$ &
				4.6 {\scriptsize(3.7--5.8)} & 3.4 {\scriptsize(2.6--4.4)} & 5.3 {\scriptsize(4.3--6.5)} &
				3.2 {\scriptsize(1.9--5.4)} & 3.6 {\scriptsize(2.2--5.9)} & 5.2 {\scriptsize(3.4--7.9)} & -- & --\\
				Fair, $\aout=0.8$ &
				15.9 {\scriptsize(13.8--18.3)} & 10.3 {\scriptsize(8.6--12.3)} & 5.3 {\scriptsize(4.1--6.9)} &
				21.0 {\scriptsize(17.3--25.3)} & 9.2 {\scriptsize(6.7--12.4)} & 9.6 {\scriptsize(7.1--12.9)} &
				6.2 {\scriptsize(4.4--8.7)} & 3.4 {\scriptsize(2.1--5.4)}\\
				Unfair, $\aout=0.8$ &
				64.6 {\scriptsize(60.3--68.7)} & 53.8 {\scriptsize(49.4--58.1)} & 73.4 {\scriptsize(69.4--77.1)} &
				68.4 {\scriptsize(63.7--72.8)} & 27.0 {\scriptsize(22.9--31.6)} & 28.4 {\scriptsize(24.2--33.0)} &
				43.8 {\scriptsize(39.5--48.2)} & 22.4 {\scriptsize(19.0--26.3)}\\
				\bottomrule
		\end{tabular}}
	\end{table*}
	
	\subsection{\fix{Baseline: without benign clustering, every audit is valid}}
	When only merit drives outcomes, every procedure rejects at or below the nominal $5\%$ in both analyses \ig{(}Table~\ref{tab:main}, first row\ig{)}. This is expected: without benign clustering the true fair probabilities are a logistic function of merit alone, so the merit-only and spatial fair worlds both contain the truth, and in the oracle analysis every reference is correctly specified. This is the world in which a single threshold was calibrated, and the result certifies the procedure there and nowhere else.

	\begingroup\fixcolor
	\subsection{Step 2: the declared fair world decides the verdict}
	\label{sec:ev-fairworld}
	\ig{This section adds benign clustering to the outcomes while keeping the system fair.} \chg{This comparison is made in the oracle analysis, where the protected context, merit and peer graph are the true ones and $\Tdec$ is the oracle statistic, so the only thing that differs between audits is the fair world from which outcomes are resampled (Table~\ref{tab:knows}, first block).} The disparity statistic of each city is the same whichever fair world it is judged against, yet the verdicts differ sharply. Judged against a fair world in which outcomes depend on merit only, which is what an analyst writes by default, 15.9\% of fair cities are declared unfair; judged against the true fair world, 5.3\% are. The whole difference lies in the declaration: a fair world with no room for legitimate clustering produces resampled cities that are too uniform, so ordinary chance alignment between the benign field and the protected geography looks extreme. \chg{The merit-only fair world is correctly specified in its merit part; its misspecification is exactly the omitted term $\aout\zout$.}
	
	The spatial fair world of Equation~\eqref{eq:fairworld} repairs this in proportion to its flexibility. \chg{It is still misspecified, but less so: a smooth surface built from $r^2$ Gaussian bumps approximates the Matérn benign field without reproducing it, and the approximation improves as $r$ grows.} Across lattice side lengths $r=2,\dots,6$ the false-rejection rate is 10.4, 4.4, 3.0, 3.2 and 0.6\%, and the detection rate at $\dd=1.25$ is 55.0, 46.6, 30.0, 14.6 and 11.0\%. The coarsest lattice is not valid. Its apparent extra sensitivity is bought with its excess false rejections: holding every lattice to an exact 5\% level, the detection rates become 46.0, 48.0, 35.0, 26.8 and 24.6\%, so $r=3$ is as sensitive as $r=2$ once both are valid. Beyond $r=3$ the loss is real, because a surface flexible enough to absorb the benign field also absorbs the part of the protected group's geography that varies on the same scale \citep{hodges2010adding,paciorek2010importance}. The choice of $r$ is therefore consequential and cannot be delegated to the data: at the same effect size, a fair world declared with $r=3$ detects unfairness in 46.6\% of unfair cities and one declared with $r=5$ in 14.6\%. We use $r=3$, fixed before any calibrated result was examined, as the operating point below\chg{; Section~\ref{sec:ev-sweep} shows that its adequacy depends on the size of the audit}.
	
	\endgroup
	\subsection{Step 5: \fix{borrowed thresholds do not travel}}
	\label{sec:ev-threshold}
	In the feasible analysis, a \fix{transported threshold (Definition~\ref{def:transport})} is set so that $5\%$ of fair cities without benign clustering exceed it, and is then applied to fair cities with benign clustering, as a regulator who published a single number would do. Concretely, for each variant the threshold $\tau$ is the 95th percentile of $\Tdec$ over 2000 simulated fair cities without clustering, computed once and never recomputed; it is a reference built from the wrong population (Table~\ref{tab:knows}, second block). It flags 21.0\% of them when the protected context and merit are known and 9.2\% when they are estimated (Table~\ref{tab:main}). \ig{The transported threshold fails more often when the protected context and merit are known than when they are estimated, which shows} that the failure lies in the threshold and not in the estimator.
	
	The mechanism is the pedestal \ig{of} Equation~\eqref{eq:pedestal}. Benign clustering barely moves its average (from 0.0424 to 0.0439), but it moves it substantially from city to city, because in each realised city the benign field and the protected geography co-vary by chance. The between-city share of the variance of the disparity statistic rises from 8\% to 67\% (Appendix~\ref{app:results}, Table~\ref{tab:ltv}). A threshold set where cities were alike is then crossed by fair cities whose pedestal happens to be high. Estimating the protected context shrinks the location shift but cannot remove the spread, because the spread belongs to the geography, not to the estimate.
	
	The calibrated reference \fix{(Definition~\ref{def:cond})} is built differently, city by city. The spatial fair world with $r=3$ is fitted to the city's own outcomes and residualised merit; 499 outcome vectors are drawn from the fitted probabilities; $\Tdec$ is recomputed for each with the 1000 posterior draws of the protected context and the peer graph held fixed; and $p_{\mathrm{cond}}$ is the share of resampled values at least as large as the observed one, as in Equation~\eqref{eq:pcond}. Its departures from the truth are the lattice approximation and, in the plugin audit, estimation of $\pi$, of merit and of $\widehat p_0$ from the data being tested. Calibrating each audit against the declared fair world fitted to its own population (Step~5) reduces the false-rejection rate under benign clustering from 9.2\% to 3.4\% (2.1--5.4) in the feasible analysis, and from 21.0\% to 6.2\% in the known-nuisance benchmark, without refitting the spatial models. It costs 18\% of detections: the feasible detection rate falls from 27.4\% to 22.4\% on the same refitted posteriors. The higher detection rate of the transported threshold was obtained at a false-rejection rate it does not control.
	\subsection{Step 3: \fix{the cost of estimating the protected context}}
	\label{sec:ev-estimation}
	In the feasible analysis the protected context is estimated by a binomial SPDE model for $A$, whose Matérn form matches the simulated surface, so its error is one of estimation, not specification. The location effects in the features are estimated by a stationary Gaussian SPDE model, which cannot represent the deterministic trends $g_k$ added to the simulated features\ig{. This} misspecification is deliberate, so that residualisation is never exact. Hyperparameters are fixed at their posterior mode. At the same effect size, an audit that knew the protected context would detect 43.8\% of unfair cities under Step~5 calibration; the feasible audit detects 22.4\%. The loss is a shrinkage of the estimated protected context towards its mean, which attenuates the disparity statistic by about a fifth. It is the price of estimating the protected group's geography from individual records, and it is the main limit on what a feasible audit can detect. It is not a reason to borrow a threshold: doing so recovers a few detections only by giving up control of false findings. Rebuilding the peer graph at every posterior draw changed at most 13 of 500 decisions per cell, and changing the spatial priors changed none detectably (Appendix~\ref{app:results}).
	\subsection{\fix{Robustness: effect size, clustering range and sample size}}
	\label{sec:ev-sweep}
	The feasible Bayesian analysis is expensive, so it was run at one configuration. The oracle analysis is not, and we repeated it, with the protected context and merit known and the same four fair worlds as in the first block of Table~\ref{tab:knows}, across the unfairness magnitude $\dd\in\{0,0.25,\dots,2\}$ with and without benign clustering, across the range of the benign field ($\rhoout\in\{0.10,0.20,0.30,0.50,0.75\}$ against $\rho_\pi=0.25$) and across city size ($n\in\{400,800,1600\}$), with 500 cities per setting and every fair world evaluated on the same resampled outcomes (Figure~\ref{fig:sweep}). Three findings bear on the protocol.
	
	\begin{figure*}[htbp]
		\centering
		\figslot{figures/fig_sweep.pdf}{6cm}
		\caption{Oracle-analysis sweeps, 500 cities per point. (a)~Rejection rate against unfairness magnitude with benign clustering ($\aout=0.8$, $\rhoout=0.30$, $n=800$); bands are 95\% Wilson intervals, and the value at $\dd=0$ is the false-rejection rate. The merit-only and $r=2$ fair worlds sit above the oracle at small $\dd$ because they are not valid, not because they are more sensitive. (b)~False-rejection rate in fair cities with benign clustering against city size, for each declared fair world and for the threshold transported from fair cities without clustering at $n=800$. (c)~Standard deviation of the disparity statistic in fair cities with benign clustering, split into its between-city part (the pedestal) and its within-city part; labels give the pedestal's share of the variance.}
		\label{fig:sweep}
	\end{figure*}
	
	\paragraph{Borrowed thresholds fail everywhere benign clustering is present, and more data does not help.} A threshold calibrated on fair cities without clustering flags 16.8 to 24.6\% of fair cities with clustering, whatever the range of the benign field, and 25.0, 21.0 and 17.8\% at $n=400$, $800$ and $1600$; without clustering it flags 4.2\%. The reason is visible in Figure~\ref{fig:sweep}(c). The within-city variation of the statistic shrinks as the city grows (standard deviation 0.0046, 0.0033, 0.0023), but the between-city variation of the pedestal does not (0.0050, 0.0047, 0.0046), because it is driven by the chance alignment of two smooth surfaces over a fixed region, which more residents sample more finely but do not average away. The pedestal's share of the variance therefore rises from 54\% to 67\% to 80\%. A larger audit measures its own city more precisely, which makes the difference between cities, and hence the error of a borrowed threshold, relatively more important. The share also rises with the range of the benign field, from 46\% at $\rhoout=0.10$ to 75\% at $\rhoout\ge0.50$.
	
	\paragraph{An under-declared fair world is exposed by larger samples.} A fair world that omits legitimate variation is a false model of the fair system, and a larger audit detects that it is false. This is the specification error of Table~\ref{tab:knows} becoming visible: the merit-only fair world omits the benign field entirely, and the lattice surface approximates it, so both are rejected more often as the data grow, while the true fair world, which is exact, is not. With benign clustering, the merit-only fair world flags 11.0, 16.4 and 21.4\% of fair cities at $n=400$, $800$ and $1600$; the spatial fair world flags 6.2, 10.4 and 17.4\% at $r=2$ and 5.0, 6.8 and 10.2\% at $r=3$; the true fair world stays at 4.4, 5.2 and 5.0\%. At $n=800$ the $r=3$ rate of 6.8\% (4.9--9.4) in this independent replication is somewhat higher than the 4.4\% of the main study; pooled over both, it is 5.6\% of 1000 fair cities. A declaration that is adequate for a small audit can therefore be inadequate for a large one. The protocol's response is to state the declaration as the spatial scale of variation treated as legitimate (the knot spacing of the lattice, which has a meaning that does not depend on $n$) and to report the verdict under the neighbouring declarations as well (Section~\ref{sec:regulators}).
	
	\paragraph{Declaring legitimacy is not free.} Without benign clustering, the spatial fair world at $r=3$ is conservative (2.2\% of fair cities flagged) and detects 47.8\% of unfair cities at $\dd=1.25$, against 78.6\% for the true fair world. Allowing spatial variation that is not there costs detection. Across the settings with clustering, once every fair world is held to an exact 5\% level, $r=2$ and $r=3$ differ by at most eight percentage points, with no consistent sign ($r=2$ ahead along the curve in $\dd$ at $n=800$, $r=3$ ahead at most other ranges and at $n=1600$); most of the coarser lattice's higher nominal detection rate in Figure~\ref{fig:sweep}(a) is bought with false findings. The operating point should therefore be declared on substantive grounds, not maximised in either direction.
	
	\fix{Rates over many cities are what a methodologist needs; a regulator sees one city at a time. Section~\ref{sec:worked} shows what the protocol delivers for a single audit.}

	\section{A worked audit report}
	\label{sec:worked}
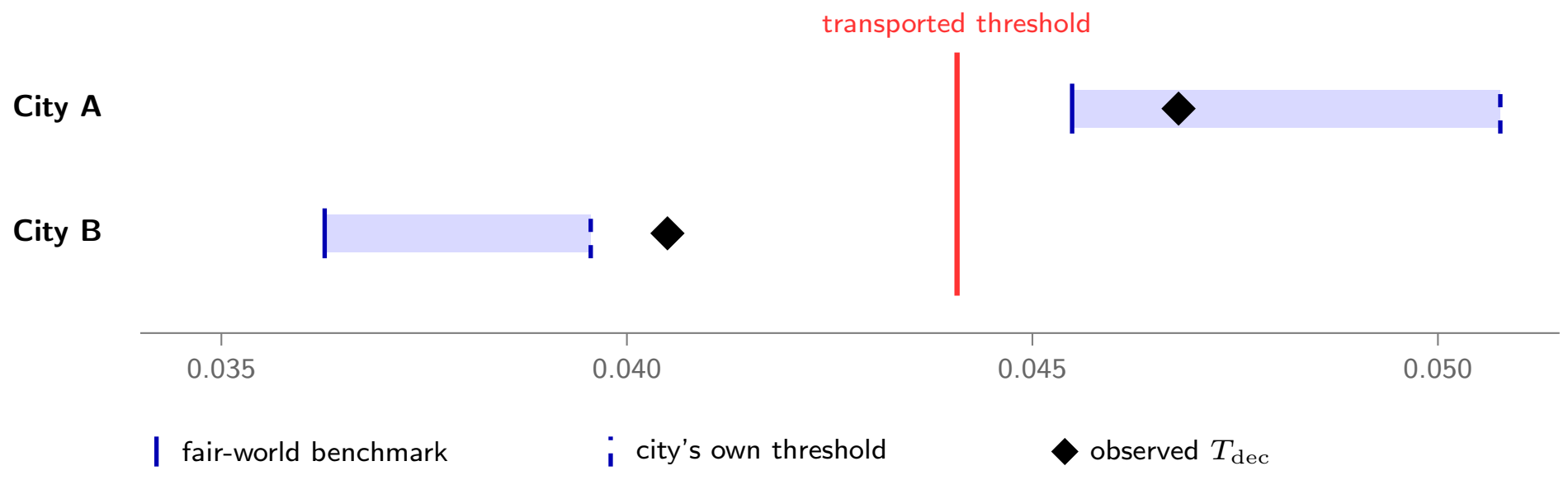
\begin{figure*}[htbp]
		\centering
		\begin{tikzpicture}[x=6.5mm, y=1cm, font=\sffamily\small]
			\draw[black!50] (0,-0.2) -- (17.5,-0.2);
			\foreach \v/\l in {1/0.035, 6/0.040, 11/0.045, 16/0.050}
			{\draw[black!50] (\v,-0.2) -- (\v,-0.3) node[below, font=\sffamily\scriptsize, text=black!60] {\l};}
			\node[anchor=east, font=\sffamily\small\bfseries] at (-0.3,0.6) {City B};
			\fill[blue!15] ({(0.03627-0.034)*1000},0.45) rectangle ({(0.03955-0.034)*1000},0.75);
			\draw[blue!70!black, line width=1pt] ({(0.03627-0.034)*1000},0.4) -- ++(0,0.4);
			\draw[blue!70!black, dashed, line width=1pt] ({(0.03955-0.034)*1000},0.4) -- ++(0,0.4);
			\node[diamond, fill=black, inner sep=2pt] at ({(0.04049-0.034)*1000},0.6) {};
			\node[anchor=east, font=\sffamily\small\bfseries] at (-0.3,1.6) {City A};
			\fill[blue!15] ({(0.04549-0.034)*1000},1.45) rectangle ({(0.05077-0.034)*1000},1.75);
			\draw[blue!70!black, line width=1pt] ({(0.04549-0.034)*1000},1.4) -- ++(0,0.4);
			\draw[blue!70!black, dashed, line width=1pt] ({(0.05077-0.034)*1000},1.4) -- ++(0,0.4);
			\node[diamond, fill=black, inner sep=2pt] at ({(0.04680-0.034)*1000},1.6) {};
			\draw[red!80, line width=1.2pt] ({(0.04407-0.034)*1000},0.1) -- ({(0.04407-0.034)*1000},2.05)
			node[above, font=\sffamily\scriptsize, text=red!80] {transported threshold};
			\begin{scope}[shift={(0.2,-1.15)}, font=\sffamily\scriptsize]
				\draw[blue!70!black, line width=1pt] (0,-0.12) -- (0,0.12); \node[anchor=west] at (0.15,0) {fair-world benchmark};
				\draw[blue!70!black, dashed, line width=1pt] (5.6,-0.12) -- (5.6,0.12); \node[anchor=west] at (5.75,0) {city's own threshold};
				\node[diamond, fill=black, inner sep=1.6pt] at (11.2,0) {}; \node[anchor=west] at (11.35,0) {observed $T_{\mathrm{dec}}$};
			\end{scope}
		\end{tikzpicture}
		\caption{The two cities of Table~\ref{tab:report} on one scale. The shaded band runs from each city's fair-world benchmark to its own 95th percentile. City~A is fair: its observed statistic lies inside the range its own fair world produces, but above the transported threshold, which would condemn it. City~B is unfair: its observed statistic lies beyond its own fair world, but below the transported threshold, which would clear it. The two cities differ in their benchmarks by more than the transported threshold can absorb.}
		\label{fig:report}
	\end{figure*}

	What does a regulator receive? Table~\ref{tab:report} shows the Step~6 report for two of the simulated cities audited by the feasible procedure at the declared operating point $r=3$, and Figure~\ref{fig:report} places their numbers on one scale. City~A is fair, with benign clustering (a B3 city). City~B is unfair (a B16 city). Both were chosen because they illustrate the central point; across all 500 cities of each kind, the rates of Table~\ref{tab:main} apply.
	
	\begin{table*}[htbp]
		\centering
		\caption{Step~6 report for two audited cities (feasible analysis, $r=3$, $B=499$ resamples). The fair-world benchmark is the mean of the decision statistic over outcome vectors regenerated from the declared fair world fitted to that city; its 95th percentile is the city's own threshold. The transported threshold is the single number calibrated on fair cities without benign clustering.}
		\label{tab:report}
		\small
		\begin{tabular}{lcc}
			\toprule
			& City A & City B\\
			\midrule
			\multicolumn{3}{l}{\emph{Declarations (Steps 1--2)}}\\
			Merit features & \multicolumn{2}{c}{three recorded features, adjusted for location}\\
			Fair world & \multicolumn{2}{c}{merit plus spatial surface, $r=3$}\\
			\midrule
			\multicolumn{3}{l}{\emph{Computation (Steps 3--5)}}\\
			Decision statistic $\Tdec$ & 0.0468 & 0.0405\\
			Fair-world benchmark (mean) & 0.0455 & 0.0363\\
			Spread of fair world (s.d.) & 0.0032 & 0.0020\\
			City's own threshold (95th percentile) & 0.0508 & 0.0396\\
			Calibrated $p$-value & 0.31 & 0.018\\
			\midrule
			\multicolumn{3}{l}{\emph{Verdicts}}\\
			Protocol (Step 5) & not flagged & \textbf{flagged}\\
			Transported threshold $\tau=0.0441$ & \textbf{flagged} & not flagged\\
			\midrule
			Truth & fair & unfair\\
			\bottomrule
		\end{tabular}
	\end{table*}
	
	The two cities make the paper's argument in a single comparison. City~A's decision statistic, 0.0468, is above the transported threshold of 0.0441, so a regulator applying a published number would declare it unfair. Yet a fair system in City~A, with its particular geography, would itself produce values around 0.0455 and up to 0.0508; the observed value is unremarkable ($p_{\mathrm{cond}}=0.31$), and the protocol does not flag it. City~B's statistic, 0.0405, is below the transported threshold, so the published number would clear it. Yet a fair system in City~B would produce values around 0.0363 and rarely above 0.0396; the observed value lies beyond that range ($p_{\mathrm{cond}}=0.018$), and the protocol flags it. The difference between the two benchmarks, 0.0092, is larger than the margin by which either city sits on the wrong side of the transported threshold. No single number can serve both cities.
	
	The report also shows what remains a judgement. Both verdicts are conditional on the declared fair world. Had the regulator declared a finer spatial surface, City~B's benchmark would rise towards its observed value and its flag could disappear. The report therefore states the declaration alongside the verdict, so that the verdict can be checked, contested and compared.
	
	\section{What a regulator should require}
	\label{sec:regulators}\label{sec:ai-statistics-governance}
	The Royal Statistical Society asks that high-stakes deployments provide evidence of performance under distribution shift and that public-sector deployments report their distributional assumptions \ig{(footnote~\ref{fn:rss})}. Our results extend both requirements to the audit itself: an audit is a statistical procedure, and a threshold carried from one population to another is an audit operating under distribution shift. A fairness audit submitted as evidence about decisions made across a geography should therefore state:
	\begin{enumerate}
		\item The fairness estimand and the merit features that define comparability, with the substantive rationale for each feature and the treatment of location-associated variation in them (Step~1);
		\item The declared fair world, including the \chg{spatial scale of variation it treats as legitimate}, fixed and recorded before the outcome data were examined, \chg{together with the verdict under the neighbouring, coarser and finer, declarations} (Step~2);
		\item Confirmation that the nuisance models were fitted without the outcomes (Step~3);
		\item the posterior distribution of the disparity, or at least its centre, spread and decision statistic (Step~4);
		\item The calibration: the fair-world benchmark for the audited population, its spread, and the calibrated $p$-value; a threshold carried over from another population should be treated as evidence under distribution shift and justified as such (Step~5);
		\item The scope of a rejection: evidence of protected-context-aligned disparity relative to the declared fair world, not proof of a causal mechanism of discrimination (Step~6).
	\end{enumerate}
	Items~1 and~2 are where regulators, domain experts and affected communities exercise judgement; items~3 to~6 are where statistical capacity is needed to check that the judgement was implemented faithfully. Recording the declarations before the data are examined, for example in a time-stamped audit plan, serves the same purpose as pre-registration in a clinical trial: it prevents the operating point from being chosen to produce a verdict.
	
	The regulatory instruments cited in Section~\ref{sec:intro} could adopt this structure without adopting \BFair{}. What they should not adopt is a fixed numerical threshold for a disparity statistic computed on spatial data. The fair benchmark of such a statistic depends on the geography of the audited population, so a fixed threshold is too strict in some places and too lenient in others, as the worked report shows. What can be standardised across audits is the procedure: who declares what, when, and against which fair world the system is calibrated.
	
	Recalibration for every audit is sometimes seen as a weakness. We see it as a consequence of what is being measured. How strongly location carries information about protected status differs between places; where it carries little, the disparity is small for most comparable pairs and the audit correctly has little to say. A threshold imported from a population in which location and protected status are more strongly associated would misstate the evidence in either direction.
	
	\section{\fix{Scope and limitations}}
	\label{sec:limits}\label{sec:discussion}
	The protocol makes the normative content of an audit explicit; it does not remove it. A rejection is relative to the declared comparability standard and fair world, and a different declaration can yield a different verdict. The estimand concerns disparity aligned with the protected group's geography among comparable individuals: outcome differences between comparable people in the same protected context, and spatial disadvantage that does not track the protected context, contribute nothing to it, although both may be unjust and both may reflect history. In the simulations, the individual protected attribute does not enter the outcome directly, so individual-level disparate treatment within a neighbourhood is not examined.
	
	\chg{The oracle analysis was repeated across effect sizes, ranges of benign clustering and sample sizes (Section~\ref{sec:ev-sweep}); the feasible Bayesian analysis was run at one configuration ($n=800$, $\dd\in\{0,1.25\}$, benign range $0.30$), and its rates elsewhere are inferred from the oracle analysis rather than measured.} The feasible calibration estimates the fair world from the same outcomes it tests, so its level is approximate; it was conservative in our study (3.4\%), and a double bootstrap or sample splitting could make it exact at additional cost \citep{hjort2006post}. Hyperparameters of the nuisance models are fixed at their posterior mode. The simulated cities treat location-associated variation in recorded features as not merit, in line with Step~1; a system whose legitimate merit is itself spatially structured is not examined. Finally, the feasible audit detects about one unfair city in five at the effect size studied, so a failure to reject is weak evidence of fairness and should be reported as such.
	
	\section{Conclusion}
	\label{sec:conclusion}
	Regulators are being asked to judge fairness evidence about algorithmic systems, and the Royal Statistical Society is right that this requires statistical capacity. For decisions about people in places, the first thing that capacity must recognise is that the benchmark against which a disparity is judged belongs to the population, not to the statistic. A fair system scores above zero, by an amount that varies with the realised geography, so a single published threshold condemns fair systems in some places and clears unfair ones in others. The protocol proposed here replaces the threshold with a procedure: declare who is comparable and which spatial variation is legitimate before seeing the outcomes, fit the spatial nuisance models without the outcomes, compute the disparity with its uncertainty, calibrate it against the declared fair world in the audited population, and report the declarations alongside the verdict. What no calibration can supply is the judgement of which spatial variation is legitimate; the protocol makes that judgement explicit and requires it to be made in advance. A fairness threshold is not a portable standard. An audit procedure can be.

	\section*{Data and code availability}
	Code (version 2026-08-24e), seeds, dataset-level output and checksums are available at\\ \ig{{\small\url{https://github.com/GandhiJafta/bfair-audit-protocol}}}. Every oracle-analysis quantity is reproducible exactly; quantities that depend on INLA posterior draws are reproducible in distribution.
	
	\bibliography{bib_fixed}

@inproceedings{dwork2012fairness,
  author    = {Dwork, Cynthia and Hardt, Moritz and Pitassi, Toniann and Reingold, Omer and Zemel, Richard},
  title     = {Fairness through awareness},
  booktitle = {Proceedings of the 3rd Innovations in Theoretical Computer Science Conference},
  pages     = {214--226},
  year      = {2012}
}

@article{mahalanobis1930tests,
  author  = {Mahalanobis, Prasanta Chandra},
  title   = {On tests and measures of group divergence},
  journal = {Journal and Proceedings of the Asiatic Society of Bengal, New Series},
  volume  = {26},
  pages   = {541--588},
  year    = {1930}
}

@article{lindgren2011explicit,
  author  = {Lindgren, Finn and Rue, H{\aa}vard and Lindstr{\"o}m, Johan},
  title   = {An explicit link between {G}aussian fields and {G}aussian {M}arkov random fields: {T}he stochastic partial differential equation approach},
  journal = {Journal of the Royal Statistical Society Series B: Statistical Methodology},
  volume  = {73},
  number  = {4},
  pages   = {423--498},
  year    = {2011}
}

@inproceedings{hardt2016equality,
  author    = {Hardt, Moritz and Price, Eric and Srebro, Nati},
  title     = {Equality of opportunity in supervised learning},
  booktitle = {Advances in Neural Information Processing Systems},
  volume    = {29},
  year      = {2016}
}

@inproceedings{feldman2015certifying,
  author    = {Feldman, Michael and Friedler, Sorelle A. and Moeller, John and Scheidegger, Carlos and Venkatasubramanian, Suresh},
  title     = {Certifying and removing disparate impact},
  booktitle = {Proceedings of the 21st ACM SIGKDD International Conference on Knowledge Discovery and Data Mining},
  pages     = {259--268},
  year      = {2015}
}

@book{openshaw1983modifiable,
  author    = {Openshaw, S.},
  title     = {The Modifiable Areal Unit Problem},
  series    = {Concepts and Techniques in Modern Geography},
  number    = {38},
  publisher = {Geo Books},
  address   = {Norwich},
  year      = {1984}
}

@inproceedings{sacharidis2023auditing,
  author    = {Sacharidis, Dimitris and Giannopoulos, Giorgos and Papastefanatos, George and Stefanidis, Kostas},
  title     = {Auditing for spatial fairness},
  booktitle = {Proceedings of the 26th International Conference on Extending Database Technology},
  year      = {2023}
}

@article{dadashpoor2025defining,
  author  = {Dadashpoor, Hashem and Dehghan, Roya},
  title   = {Defining spatial justice: {A} review},
  journal = {Habitat International},
  volume  = {160},
  pages   = {103387},
  year    = {2025}
}

@inproceedings{xie2022fairness,
  author    = {Xie, Yiqun and He, Erhu and Jia, Xiaowei and Chen, Weiye and Skakun, Sergii and Bao, Han and Jiang, Zhe and Ghosh, Rahul and Ravirathinam, Praveen},
  title     = {Fairness by ``where'': {A} statistically-robust and model-agnostic bi-level learning framework},
  booktitle = {Proceedings of the Thirty-Sixth AAAI Conference on Artificial Intelligence},
  volume    = {36},
  pages     = {12208--12216},
  publisher = {AAAI Press},
  year      = {2022}
}

@inproceedings{saxena2025legally,
  author    = {Saxena, Nripsuta Ani and Mathur, Ronit and Shahabi, Cyrus},
  title     = {Legally-compliant spatial fairness framework: {A}dvancing beyond spatial fairness},
  booktitle = {Proceedings of the 28th International Conference on Extending Database Technology},
  year      = {2025}
}

@article{shaham2022models,
  author  = {Shaham, Sina and Ghinita, Gabriel and Shahabi, Cyrus},
  title   = {Models and mechanisms for spatial data fairness},
  journal = {Proceedings of the Very Large Data Bases Endowment},
  volume  = {16},
  number  = {2},
  pages   = {167--179},
  year    = {2022}
}

@article{baresi2025neo,
  author  = {Baresi, Umberto},
  title   = {Neo-colonial intelligence: {H}ow {AI} risks reinforcing spatial injustices in a digitally divided world},
  journal = {Cities},
  volume  = {166},
  pages   = {106232},
  year    = {2025}
}

@article{dupont2022spatial,
  author  = {Dupont, Emiko and Wood, Simon N. and Augustin, Nicole H.},
  title   = {Spatial+: {A} novel approach to spatial confounding},
  journal = {Biometrics},
  volume  = {78},
  number  = {4},
  pages   = {1279--1290},
  year    = {2022}
}

@article{rue2009inla,
  author  = {Rue, H{\aa}vard and Martino, Sara and Chopin, Nicolas},
  title   = {Approximate {B}ayesian inference for latent {G}aussian models by using integrated nested {L}aplace approximations},
  journal = {Journal of the Royal Statistical Society Series B: Statistical Methodology},
  volume  = {71},
  number  = {2},
  pages   = {319--392},
  year    = {2009}
}

@inproceedings{dimitrakakis2019bayesian,
  author    = {Dimitrakakis, Christos and Liu, Yang and Parkes, David C. and Radanovic, Goran},
  title     = {Bayesian fairness},
  booktitle = {Proceedings of the Thirty-Third AAAI Conference on Artificial Intelligence},
  volume    = {33},
  year      = {2019}
}

@article{hodges2010adding,
  author  = {Hodges, James S. and Reich, Brian J.},
  title   = {Adding spatially-correlated errors can mess up the fixed effect you love},
  journal = {The American Statistician},
  volume  = {64},
  number  = {4},
  pages   = {325--334},
  year    = {2010}
}

@article{barocas2016big,
  author  = {Barocas, Solon and Selbst, Andrew D.},
  title   = {Big data's disparate impact},
  journal = {California Law Review},
  volume  = {104},
  number  = {3},
  pages   = {671--732},
  year    = {2016}
}

@inproceedings{barrainkua2024uncertainty,
  author    = {Barrainkua, Ainhize and Gordaliza, Paula and Lozano, Jose A. and Quadrianto, Novi},
  title     = {Uncertainty matters: {S}table conclusions under unstable assessment of fairness results},
  booktitle = {Proceedings of the 27th International Conference on Artificial Intelligence and Statistics},
  series    = {Proceedings of Machine Learning Research},
  volume    = {238},
  pages     = {1198--1206},
  publisher = {PMLR},
  year      = {2024}
}

@article{van2023new,
	title={{A new avenue for Bayesian inference with INLA}},
	author={Van Niekerk, Janet and Krainski, Elias and Rustand, Denis and Rue, H{\aa}vard},
	journal={{Computational Statistics \& Data Analysis}},
	volume={181},
	pages={107692},
	year={2023},
	publisher={Elsevier}
}

@article{simpson2017penalising,
  author  = {Simpson, Daniel and Rue, H{\aa}vard and Riebler, Andrea and Martins, Thiago G. and S{\o}rbye, Sigrunn H.},
  title   = {Penalising model component complexity: {A} principled, practical approach to constructing priors},
  journal = {Statistical Science},
  volume  = {32},
  number  = {1},
  pages   = {1--28},
  year    = {2017}
}

@inproceedings{saxena2024spatial,
  author    = {Saxena, Nripsuta Ani and Horn, Abigail L. and Zhang, Wenbin and Shahabi, Cyrus},
  title     = {Position: {S}patial fairness: {F}oundations, pitfalls, and a path forward},
  booktitle = {Proceedings of the 43rd International Conference on Machine Learning},
  series    = {Proceedings of Machine Learning Research},
  year      = {2026}
}

@inproceedings{shaham2024fairkd,
  author    = {Shaham, Sina and Ghinita, Gabriel and Shahabi, Cyrus},
  title     = {Fair spatial indexing: {A} paradigm for group spatial fairness},
  booktitle = {Proceedings of the 27th International Conference on Extending Database Technology},
  pages     = {150--161},
  year      = {2024}
}

@inproceedings{he2022sailing,
  author    = {He, Erhu and Xie, Yiqun and Jia, Xiaowei and Chen, Weiye and Bao, Han and Zhou, Xun and Jiang, Zhe and Ghosh, Rahul and Ravirathinam, Praveen},
  title     = {Sailing in the location-based fairness-bias sphere},
  booktitle = {Proceedings of the 30th International Conference on Advances in Geographic Information Systems},
  pages     = {1--10},
  publisher = {ACM},
  year      = {2022}
}

@inproceedings{jafta2021evaluation,
  author    = {Jafta, Gandhi and de Waal, Alta and Derks, Iena and Ruttkamp-Bloem, Emma},
  title     = {Evaluation of {XAI} as an enabler for fairness, accountability and transparency},
  booktitle = {Artificial Intelligence Research: Second Southern African Conference, SACAIR 2021, Durban, South Africa, December 6--10, 2021, Proceedings},
  series    = {Communications in Computer and Information Science},
  volume    = {1551},
  publisher = {Springer},
  address   = {Cham},
  year      = {2021}
}

@article{binns2018algorithmic,
  author  = {Binns, Reuben},
  title   = {Algorithmic accountability and public reason},
  journal = {Philosophy \& Technology},
  volume  = {31},
  number  = {4},
  pages   = {543--556},
  year    = {2018}
}

@unpublished{jafta2026bfair,
  author = {Jafta, Gandhi and Fabris-Rotelli, Inger and Ruttkamp-Bloem, Emma and van Niekerk, Janet and Maharela, Iketle},
  title  = {{B-Fair}: {A} {B}ayesian spatial individual fairness framework},
  note   = {Manuscript under review},
  year   = {2026}
}

@article{Giudici09072026,
  author  = {Giudici, Paolo and Neelakantan, Parvati and Pavarana, Simone and Schmidt, Thorsten},
  title   = {Are bank decisions fair? {A} statistical analysis of fairness in credit lending},
  journal = {Statistics: A Journal of Theoretical and Applied Statistics},
  year    = {2026}
}

@article{kunsch1989jackknife,
  author  = {K{\"u}nsch, Hans R.},
  title   = {The jackknife and the bootstrap for general stationary observations},
  journal = {The Annals of Statistics},
  volume  = {17},
  number  = {3},
  pages   = {1217--1241},
  year    = {1989}
}

@book{lahiri2003resampling,
  author    = {Lahiri, Soumendra Nath},
  title     = {Resampling Methods for Dependent Data},
  publisher = {Springer},
  address   = {New York},
  year      = {2003}
}

@book{ruppert2003semiparametric,
  author    = {Ruppert, David and Wand, Matt P. and Carroll, Raymond J.},
  title     = {Semiparametric Regression},
  publisher = {Cambridge University Press},
  year      = {2003}
}

@book{golub2013matrix,
  author    = {Golub, Gene H. and Van Loan, Charles F.},
  title     = {Matrix Computations},
  edition   = {4th},
  publisher = {Johns Hopkins University Press},
  year      = {2013}
}

@book{wood2017gam,
  author    = {Wood, Simon N.},
  title     = {Generalized Additive Models: An Introduction with {R}},
  edition   = {2nd},
  publisher = {CRC Press},
  year      = {2017}
}

@book{hastie2009elements,
  author    = {Hastie, Trevor and Tibshirani, Robert and Friedman, Jerome},
  title     = {The Elements of Statistical Learning: Data Mining, Inference, and Prediction},
  edition   = {2nd},
  publisher = {Springer},
  year      = {2009}
}

@article{cherian2024statistical,
  author  = {Cherian, John J. and Cand{\`e}s, Emmanuel J.},
  title   = {Statistical inference for fairness auditing},
  journal = {Journal of Machine Learning Research},
  volume  = {25},
  number  = {149},
  pages   = {1--49},
  year    = {2024}
}

@inproceedings{diciccio2020evaluating,
  author    = {DiCiccio, Cyrus and Vasudevan, Sriram and Basu, Kinjal and Kenthapadi, Krishnaram and Agarwal, Deepak},
  title     = {Evaluating fairness using permutation tests},
  booktitle = {Proceedings of the 26th ACM SIGKDD International Conference on Knowledge Discovery and Data Mining},
  pages     = {1467--1477},
  year      = {2020}
}

@article{fuglstad2019constructing,
  author  = {Fuglstad, Geir-Arne and Simpson, Daniel and Lindgren, Finn and Rue, H{\aa}vard},
  title   = {Constructing priors that penalize the complexity of {G}aussian random fields},
  journal = {Journal of the American Statistical Association},
  volume  = {114},
  number  = {525},
  pages   = {445--452},
  year    = {2019}
}

@article{rubin1984bayesianly,
  author  = {Rubin, Donald B.},
  title   = {Bayesianly justifiable and relevant frequency calculations for the applied statistician},
  journal = {The Annals of Statistics},
  volume  = {12},
  number  = {4},
  pages   = {1151--1172},
  year    = {1984}
}

@article{reich2006effects,
  author  = {Reich, Brian J. and Hodges, James S. and Zadnik, Vesna},
  title   = {Effects of residual smoothing on the posterior of the fixed effects in disease-mapping models},
  journal = {Biometrics},
  volume  = {62},
  number  = {4},
  pages   = {1197--1206},
  year    = {2006}
}

@article{hanks2015restricted,
  author  = {Hanks, Ephraim M. and Schliep, Erin M. and Hooten, Mevin B. and Hoeting, Jennifer A.},
  title   = {Restricted spatial regression in practice: {G}eostatistical models, confounding, and robustness under model misspecification},
  journal = {Environmetrics},
  volume  = {26},
  number  = {4},
  pages   = {243--254},
  year    = {2015}
}

@article{khan2022restricted,
  author  = {Khan, Kori and Calder, Catherine A.},
  title   = {Restricted spatial regression methods: {I}mplications for inference},
  journal = {Journal of the American Statistical Association},
  volume  = {117},
  number  = {537},
  pages   = {482--494},
  year    = {2022}
}

@article{paciorek2010importance,
  author  = {Paciorek, Christopher J.},
  title   = {The importance of scale for spatial-confounding bias and precision of spatial regression estimators},
  journal = {Statistical Science},
  volume  = {25},
  number  = {1},
  pages   = {107--125},
  year    = {2010}
}

@article{mahalanobis1936generalised,
  author  = {Mahalanobis, Prasanta Chandra},
  title   = {On the generalised distance in statistics},
  journal = {Proceedings of the National Institute of Sciences of India},
  volume  = {2},
  number  = {1},
  pages   = {49--55},
  year    = {1936}
}

@inproceedings{maity2021statistical,
  author    = {Maity, Subha and Xue, Songkai and Yurochkin, Mikhail and Sun, Yuekai},
  title     = {Statistical inference for individual fairness},
  booktitle = {International Conference on Learning Representations},
  year      = {2021}
}

@inproceedings{foulds2020bayesian,
  author    = {Foulds, James R. and Islam, Rashidul and Keya, Kamrun Naher and Pan, Shimei},
  title     = {Bayesian modeling of intersectional fairness: {T}he variance of bias},
  booktitle = {Proceedings of the 2020 SIAM International Conference on Data Mining},
  pages     = {424--432},
  year      = {2020}
}

@inproceedings{ji2020can,
  author    = {Ji, Disi and Smyth, Padhraic and Steyvers, Mark},
  title     = {Can {I} trust my fairness metric? {A}ssessing fairness with unlabeled data and {B}ayesian inference},
  booktitle = {Advances in Neural Information Processing Systems},
  volume    = {33},
  year      = {2020}
}

@article{hjort2006post,
  author  = {Hjort, Nils Lid and Dahl, Fredrik A. and Steinbakk, Gunnhildur H{\o}gn{\aa}sen},
  title   = {Post-processing posterior predictive p values},
  journal = {Journal of the American Statistical Association},
  volume  = {101},
  number  = {475},
  pages   = {1157--1174},
  year    = {2006}
}

@misc{euaiact2024,
  author       = {{European Parliament and Council of the European Union}},
  title        = {Regulation ({EU}) 2024/1689 laying down harmonised rules on artificial intelligence ({A}rtificial {I}ntelligence {A}ct)},
  howpublished = {Official Journal of the European Union, L series, 12 July 2024},
  year         = {2024}
}

@phdthesis{jafta2026calibrating,
  author = {Jafta, Gandhi},
  title  = {Bayesian individual spatial fairness},
  school = {University of Pretoria},
  type   = {{PhD} thesis in Mathematical Statistics},
  year   = {2026}
}

@book{matern1986spatial,
  author    = {Mat{\'e}rn, Bertil},
  title     = {Spatial Variation},
  edition   = {2nd},
  series    = {Lecture Notes in Statistics},
  volume    = {36},
  publisher = {Springer},
  address   = {Berlin},
  year      = {1986}
}

@book{stein1999interpolation,
  author    = {Stein, Michael L.},
  title     = {Interpolation of Spatial Data: Some Theory for Kriging},
  publisher = {Springer},
  address   = {New York},
  year      = {1999}
}
	
	\clearpage
	\raggedbottom 
	\section*{Appendix}
	\setcounter{section}{0}
	\renewcommand{\thesection}{A\arabic{section}}
	\renewcommand{\thesubsection}{\thesection.\arabic{subsection}}
	\setcounter{figure}{0}\renewcommand{\thefigure}{A\arabic{figure}}
	\setcounter{table}{0}\renewcommand{\thetable}{A\arabic{table}}
	\setcounter{equation}{0}\renewcommand{\theequation}{A\arabic{equation}}
	\section{The \BFair{} construction and its decision layer}
	\label{app:bfair}
	
	The process begins with an observed dataset
	\[
	D_n=\{(f_i,O_i,s_i,A_i)\}_{i=1}^{n},
	\]
	where, for each individual $i$,
	\begin{itemize}
		\item $s_i \in S \subset \mathbb{R}^2$ is the geographic location (coordinates);
		\item $f_i \in \mathbb{R}^q$ is the vector of $q$ raw merit features (e.g.\ income or risk scores);
		\item $O_i \in \{0,1\}$ is the algorithmic outcome, here a binary outcome with $O_i=1$ denoting the adverse outcome, so that $O_i\mid p_i\sim\Bern(p_i)$ with $p_i$ the probability of the adverse outcome;
		\item $A_i \in \{0, 1\}$ is the observed protected attribute (e.g.\ minority status).
	\end{itemize}
	The fair-null counterpart of $p_i$, and its fitted value, are introduced with the reference distribution in Section~\ref{sec:reference-calibration}.
	\subsection{Spatial latent modelling}
	The framework fits two parallel latent Gaussian models to account for geographic dependencies.
	\begin{itemize}
		\item The Merit Feature Model: Each raw feature is decomposed as $f_i = \mu + \xi(s_i) + \varepsilon_i$, where $\mu$ is the global intercept, $\xi(s_i)$ is a zero-mean spatial Mat\'{e}rn field, and $\varepsilon_i \sim N(0, \sigma^2_\varepsilon)$ is random noise.
		\item The Protected-Context Model: The protected attribute is modelled as a continuous probabilistic demographic surface $\pi(s_i)$. The log-odds are defined as $\text{logit}(\pi(s_i)) = \beta_0 + u_A(s_i)$, where $\beta_0$ is an intercept and $u_A(\cdot)$ is a zero-mean Gaussian random field with Mat\'ern covariance, whose practical range $\rho_A$ and marginal standard deviation $\sigma_A$ are hyperparameters given penalised-complexity priors (Section~\ref{sec:structure}). 

	\end{itemize}
	\subsection{The first pushforward: constructing the peer graph}
	The framework draws $M$ joint posterior samples (indexed by $m$) from these models to build relational structures:
	\begin{itemize}
		\item Spatial Residualisation: merit features are calculated as
		\[
		\tilde{f}_i^{(m)} = f_i - \bigl\{\mu^{(m)} + \xi^{(m)}(s_i)\bigr\},
		\]
		where $\mu^{(m)}$ and $\xi^{(m)}(s_i)$ are the $m$th posterior draws of the intercept and of the spatial field of the merit feature model, evaluated at individual $i$'s location. Subtracting the fitted location effect from the recorded feature isolates individual merit from geographically elicited advantage; for $q$ features the subtraction is applied feature by feature.
		\item Similarity Geometry: Similarity is measured using the Mahalanobis distance \citep{mahalanobis1936generalised}:
		
		\begin{equation} 
			d_{\tilde{f}}^{(m)}(i, j) = \sqrt{(\tilde{f}_i^{(m)} - \tilde{f}_j^{(m)})^\top(\Sigma^{(m)})^{-1}(\tilde{f}_i^{(m)} - \tilde{f}_j^{(m)})} ,
		\end{equation}
		
		where $\Sigma^{(m)} = \mathrm{Cov}(\widetilde f^{(m)}) \in \mathbb{R}^{q \times q}$ is the empirical covariance matrix of the residualised features across all $n$ individuals for posterior draw $m$. 
		The empirical covariance matrix is estimated across the $n$ residualised feature vectors within the posterior draw $m$,
		\begin{equation}
			\Sigma^{(m)} = \frac{1}{n-1} \sum_{i=1}^n \left(\widetilde f_i^{(m)} - \bar{\widetilde f}^{(m)}\right)\left(\widetilde f_i^{(m)} - \bar{\widetilde f}^{(m)}\right)^{\top},
		\end{equation}
		where $\bar{\widetilde f}^{(m)} = \frac{1}{n}\sum_{i=1}^n \widetilde f_i^{(m)}$ is the sample mean vector for draw $m$. This covariance structure reflects the joint dependence among merit features after spatial residualisation, capturing correlations that arise from socioeconomic mechanisms rather than geographic proximity. A small ridge is added to $\Sigma^{(m)}$ before inversion for numerical stability. Recomputing $\Sigma^{(m)}$ at every draw is required only in the \emph{full} variant, which rebuilds the peer graph per draw; the \emph{plugin} variant computes a single $\Sigma$ from the posterior-mean residualised features and keeps it, and the graph, fixed across draws. Section~\ref{sec:sim-results} shows that the two variants differ in at most 13 of 500 decisions per scenario, so holding the metric fixed is an adequate default. The calculated similarity distances are converted into continuous edge membership weights using a Gaussian radial basis function (RBF) kernel,
		\begin{equation}
			w_{\mathcal P_h}^{(m)}(i, j) = \exp \left( - \frac{\left[d_{\tilde{f}}^{(m)}(i, j)\right]^2}{2h^2} \right) 
		\end{equation}
		where $h>0$ is the bandwidth and the subscript $\mathcal P_h$ denotes the peer relation induced by this kernel at bandwidth $h$. A pair at distance $h$ receives weight $e^{-1/2}\approx0.61$, and weights decay towards zero for pairs that are far apart in residualised merit.
	\end{itemize}
	For each posterior draw, the peer graph is first constructed as a directed \(k\)-nearest-neighbour graph in the residualised-feature space; we use \(k=20\) throughout, a tuning choice rather than a property of the method. The directed relation is then symmetrised by union: an unordered pair \(\{i,j\}\) is retained if either \(i\) lists \(j\) or \(j\) lists \(i\) among its \(k\) nearest neighbours. Each retained unordered pair is stored once, with \(i<j\).
	Let \(d_e\) denote the Mahalanobis distance of retained peer-graph edge \(e\in\mathcal E\). The RBF bandwidth can then be defined as
	\[
	h
	=
	Q_{0.50}\{d_e:e\in\mathcal E\},
	\]
	the empirical median of the retained edge distances. The bandwidth adapts to the observed scale of substantively comparable peers but is fixed within an audit and throughout its reference resampling; for consistency, it is not retuned after observing the audit outcome. 
	
	\subsection{The second pushforward: the \BFair{} functional}
	For outcomes coded so that \(O_i=1\) is adverse, the pairwise violation
	magnitude is the hinge function
	\[
	V_\eta(O_i,O_j)
	=
	\max\{0,\lvert O_i-O_j\rvert-\eta\},
	\]
	where \(\eta\ge0\) is an indifference tolerance. For a binary outcome, $V_\eta=(1-\eta)\,\mathbf 1\{O_i\neq O_j\}$ when $\eta<1$, so $\eta$ only rescales the functional and we set \(\eta=0\) throughout. A positive tolerance matters for graded or continuous outcomes, such as scores or premiums, where differences smaller than $\eta$ are treated as immaterial. Directionality is imposed
	separately through an indicator that retains a discordant pair only when the
	individual in the higher-protected-context setting receives the adverse
	outcome. The directional version of \BFair{} uses the sign indicator
	\[
	\begin{aligned}
	\Delta_{ij}={}&\mathbb{I}\bigl\{\pi^{(m)}(s_i) > \pi^{(m)}(s_j),\, O_i > O_j\bigr\}\\
	&+ \mathbb{I}\bigl\{\pi^{(m)}(s_j) > \pi^{(m)}(s_i),\, O_j > O_i\bigr\},
	\end{aligned}
	\]
	which equals $1$ precisely when the individual in the higher-protected-context setting receives the adverse outcome.
	
	\subsection{Reference distributions}
	
	An audit needs a yardstick. A disparity score means little until we know what scores a fair system would typically produce in the same population. The reference distribution supplies that yardstick, and this section shows that its choice is a substantive decision rather than a technicality. The observed directional disparity functional should not be interpreted against an arbitrary absolute benchmark such as zero. Instead, \BFair{} calibrates an audit decision against a design-specific reference distribution. Let
	\[
	\mathcal D_{\mathrm{fixed}}
	=
	\left\{
	\widetilde f_i,s_i,\pi(s_i)
	\right\}_{i=1}^n
	\]
	denote the realised design quantities of the functional, treated as fixed within a reference replicate. The reference distribution is
	\[
	\mathcal R_0
	=
	\mathcal L\left(
	T_{\mathrm{dir}}^\star
	\mid
	H_0,\mathcal D_{\mathrm{fixed}}
	\right),
	\]
	where \(T_{\mathrm{dir}}^\star\) is evaluated after regenerating the outcome vector \(O^\star\) under a specified  substantively fair null mechanism. The reference distribution is therefore not an alternative estimate of the observed disparity. On a fixed audit dataset, alternative reference constructions leave \(\widehat T_{\mathrm{dir}}\) unchanged; they differ in the fair-world variation against which that observed value is judged. An observed value constitutes evidence against the specified fair-null reference	only when it is extreme relative to \(\mathcal R_0\). A rejection is evidence of protected-context-aligned disparity under the documented audit assumptions; it is not, by itself, causal proof of discrimination.
	
	$H_0$ in $\mathcal R_0$ is not unique: the simulation study evaluates it under several constructions, which encode different substantive fair-null hypotheses rather than different implementations of one idea. Table~\ref{tab:reference-distributions} summarises the three constructions discussed in the main text; the complete set of six, including the two used only as negative controls in the exploratory study, is given in the Supplementary Material (Table~S1).
	\begin{table*}[htbp]
		\centering
		\caption{Fair-null reference distributions discussed in the main text. All three regenerate only the outcome vector $O$, holding $\mathcal D_{\mathrm{fixed}}$ fixed within a replicate; they differ in the class of outcome mechanisms declared fair, not in the functional being tested.}
		\label{tab:reference-distributions}
		\small
		
		\begin{tabular}{lp{5.0cm}p{5.2cm}}
			\toprule
			Reference & Null hypothesis encoded & Role \\
			\midrule
			\texttt{restricted\_boot} (merit-only restricted) & $\logit p_i=\gamma_0+\widetilde f_i^\top\boldsymbol\gamma$; outcomes depend on residualised merit only & the construction an analyst writes by default; the cautionary case that over-rejects under residual spatial-outcome dependence \\
			\texttt{restricted\_spatial} & $\logit p_i=\gamma_0+\widetilde f_i^\top\boldsymbol\gamma+\widetilde{\mathbf B}_r(s_i)^\top\boldsymbol\zeta_r$; legitimate spatial variation confined to the span of the lattice basis, with no $\pi$, $A$ or proxies & deployable; frontier in lattice rank $r$ evaluated by Rule D1 \\
			\texttt{oracle\_boot} (known-nuisance benchmark) & $O_i^\star \sim \mathrm{Bernoulli}(p_i^{(0)})$, true simulated DGP & infeasible; scientific benchmark, not itself a candidate for deployment \\
			\bottomrule
		\end{tabular}
		
	\end{table*}
	
	This restricted-spatial reference is an inferential extension of the \BFair{} framework. It is not required to define the posterior directional functional itself. Rather, it specifies the fair-world variation against which an observed \BFair{} disparity is judged when the framework is used as a formal audit.
	Its spatial complexity is controlled by an integer lattice side length $r\ge2$, referred to below as the spatial-lattice side length for brevity. Importantly, $r$ is not the number of basis functions: it specifies an $r\times r$ deterministic knot lattice and hence produces up to $r^2$ Gaussian radial-basis columns before numerical rank reduction \citep{ruppert2003semiparametric, hastie2009elements}.
	
	Let $s_i=(x_i,y_i)$ denote the observed coordinate of individual $i$, and
	let
	\[
	\mathcal C_r
	=
	\{c_1,\ldots,c_{r^2}\}
	\]
	be the regular $r\times r$ knot lattice spanning the observed coordinate
	ranges. Define the knot spacing and basis bandwidth by
	\[
	\Delta_r
	=
	\frac{
		\max\{\operatorname{range}(x),\operatorname{range}(y)\}
	}{r-1},
	\qquad
	b_r=\Delta_r.
	\]
	For knot $c_\ell$, the Gaussian radial-basis column is
	\[
	B_{i\ell}^{(r)}
	=
	\exp\left\{
	-\frac{\lVert s_i-c_\ell\rVert_2^2}{2b_r^2}
	\right\},
	\qquad
	\ell=1,\ldots,r^2.
	\]
	The columns of $\mathbf B^{(r)}$ are centred and reduced to full numerical column rank using a pivoted QR decomposition \citep{golub2013matrix, wood2017gam}\footnote{Gaussian radial basis functions placed on a regular lattice can be nearly collinear, and their sum can be approximately constant, creating near-aliasing with the intercept \citep{wood2017gam}. The pivoted QR decomposition retains a full-rank subset of centred basis columns under a numerical tolerance of $10^{-7}$ \citep{golub2013matrix}. This prevents aliased binomial-regression coefficients from generating undefined fitted probabilities.}. The resulting coordinate basis $\widetilde{\mathbf B}_r(s_i)$ is deterministic conditional on the observed coordinates: knot placement and basis bandwidth do not vary across bootstrap resamples. For a selected lattice side length $r$, the restricted-spatial fair-null outcome model is
	\[
	\operatorname{logit}\left\{
	p_i^{(0,r)}
	\right\}
	=
	\gamma_0
	+
	\widetilde f_i^\top\boldsymbol{\gamma}
	+
	\widetilde{\mathbf B}_r(s_i)^\top\boldsymbol{\zeta}_r.
	\]
	The protected-context surface $\pi(s_i)$, realised protected attribute $A_i$, and direct proxies for them are excluded from this null model. \chg{Conditional on the fitted fair-null} probabilities,
	\[
	O_i^\star
	\mid
	\widehat p_i^{(0,r)}
	\sim
	\operatorname{Bernoulli}
	\left\{
	\widehat p_i^{(0,r)}
	\right\}.
	\]
	Increasing $r$ reduces knot spacing and permits the fair-null model to represent progressively finer residual spatial outcome variation \citep{ruppert2003semiparametric}. In the examined simulation family, this induces a validity--sensitivity frontier: coarse lattices can leave nuisance spatial dependence outside the null and yield excess rejection, whereas more flexible lattices can absorb spatial variation overlapping with the protected-context-aligned alternative and reduce sensitivity\ig{, a form of spatial confounding} \citep{hodges2010adding,paciorek2010importance,dupont2022spatial}.
	
	A natural alternative is to orthogonalise the spatial basis against the estimated protected context, as in restricted spatial regression \citep{reich2006effects,hodges2010adding}. This would preserve sensitivity by construction, but it would also remove from the null exactly the component of the benign field that co-moves with $\pi$ by chance in the realised population, and so reintroduce the over-rejection that the spatial reference is meant to prevent. Restricted spatial regression is moreover known to understate uncertainty and to yield anti-conservative inference \citep{hanks2015restricted,khan2022restricted}. We therefore treat the unrestricted lattice as the deployable reference and its side length as the declared operating point.
	
	\subsection{Bayesian decision rule}
	\label{sec:decision-rule}
	
	The inferential engine of the framework is the posterior distribution
	\begin{equation}
		p(T_{\mathrm{dir}}\mid\mathcal{D}),
	\end{equation}
	obtained by propagating joint posterior uncertainty about the protected-context surface and the peer-matching structure through the directional fairness functional. For a posterior tail level $\alpha\in(0,1/2)$, the decision statistic is the lower $100\alpha\%$ posterior quantile,
	\begin{equation}
		T_{\mathrm{dec}} = Q^{+}_{\alpha} \left\{ T_{\mathrm{dir}}\mid\mathcal{D} \right\},
		\qquad Q^{+}_{\alpha}=\inf\{t:F(t\mid\mathcal D)>\alpha\},
	\end{equation}
	where $F(\cdot\mid\mathcal D)$ is the posterior distribution function; the quantile $Q^{+}$ is what makes the equivalence below hold with a non-strict inequality. Given a threshold $\tau$, calibrated as described in Sections~\ref{sec:sim-implications} and~\ref{sec:conditional}, the audit rejects the specified fair null when
	\begin{equation}
		T_{\mathrm{dec}} > \tau.
		\label{eq:rule}
	\end{equation}
	For the empirical posterior of $M$ draws with $M\alpha$ an integer, $\Tdec$ is the $(M\alpha+1)$th smallest draw, and counting draws shows that
	\begin{equation}
		\Tdec>\tau
		\iff
		\widehat{\Pr}(T_{\mathrm{dir}}>\tau\mid\mathcal D)\ge1-\alpha
		\iff
		\widehat{\Pr}(T_{\mathrm{dir}}\le\tau\mid\mathcal D)\le\alpha .
		\label{eq:rule-prob}
	\end{equation}
	In words, a system is declared unfair when, under the fitted model, the disparity exceeds the calibrated threshold with posterior probability at least $1-\alpha$. We use $\alpha=0.05$ and $M=1000$ throughout, so $\Tdec$ is the 51st smallest of the 1000 draws. The posterior tail level $\alpha$ should not be confused with the audit's false-rejection rate, which is controlled by the calibration of $\tau$, and which is also $5\%$ in this paper.
	
	The lower quantile is a choice of test statistic, not an additional layer of conservatism. Once $\tau$ is set to the $95\%$ point of the sampling distribution of $\Tdec$ under a fair mechanism, the audit has level $5\%$ under that mechanism whichever posterior summary is used, provided $\tau$ is calibrated on the same summary that the rule compares with it. What the lower quantile changes is power. Because it lies below the centre of the posterior by an amount proportional to the posterior spread, an audit with a diffuse posterior must exhibit a larger disparity to be flagged than one with a concentrated posterior. The rule therefore flags a system only when the disparity is large relative to its own posterior uncertainty, and the repeated-design calibration of $\tau$ is what gives that Bayesian statement a frequency interpretation \citep{rubin1984bayesianly}. Posterior means and credible intervals are reported as descriptive summaries of effect size.

	\paragraph{Step 3: Fit the nuisance models without the outcomes}
	The protected context (the probability that a resident of each location belongs to the protected group) and the location effects in the merit features are estimated from the protected attribute, the features, and the locations, without using the outcomes under audit. These quantities define the residualised features, the peer graph, and, ultimately, the statistic applied to each outcome vector. Keeping their estimation outcome-free allows them to be held fixed while only the outcome vector is regenerated in Step~5. This is the condition needed for the exchangeability argument of Lemma~\ref{lem:cond} when the fair-null probabilities are known. It is important to distinguish these statistic-defining nuisance models from the fair-null outcome model used in Step~5: the latter is fitted to the observed outcomes and therefore produces a plug-in calibration whose finite-sample level is approximate rather than exact.
	
	\begin{figure*}[tp]
		\centering
		\includegraphics[width=0.85\textwidth]{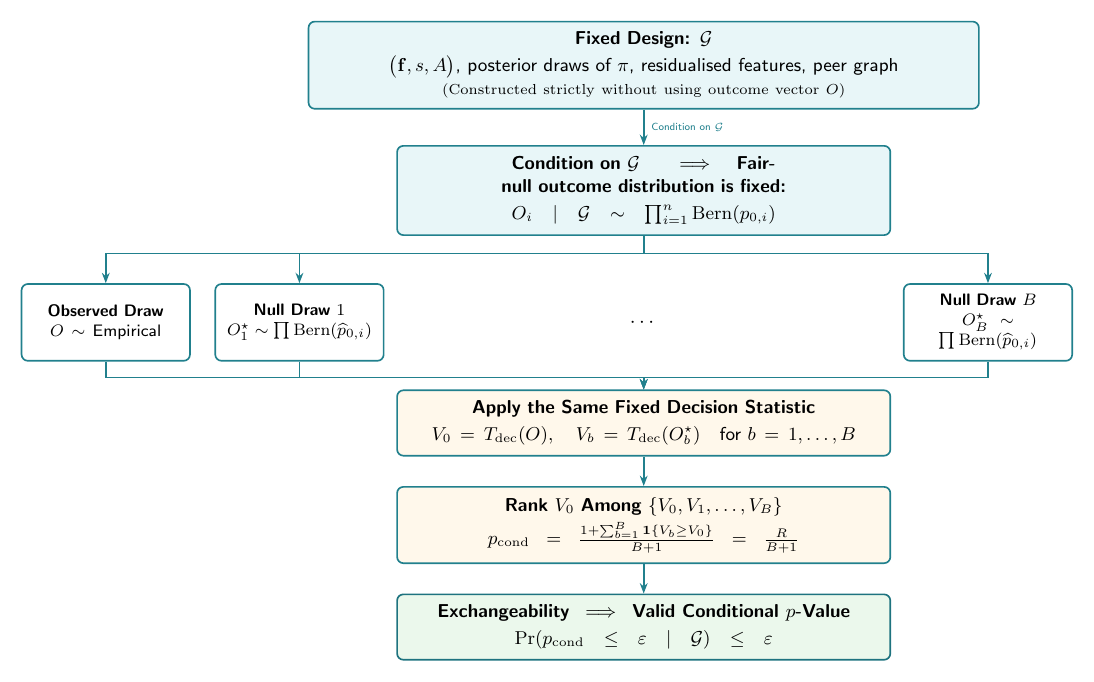}
		\caption{Conditional Monte Carlo exchangeability setup under Lemma~\ref{lem:cond}.  The nuisance parameters and peer graph in $\mathcal{G}$ are fitted without observing the outcome $O$, conditioning on $\mathcal{G}$ freezes the decision statistic $T_{\mathrm{dec}}(\cdot)$ while preserving identical, independent null outcome distributions across all $B+1$ vectors.}
		\label{fig:mc_exchangeability}
	\end{figure*}
	\begin{lemma}[Design-conditional validity of the Monte Carlo test]
		\label{lem:cond}
		Let
		\[
		\mathcal{G}
		=
		\bigl(f, s, A, \text{posterior draws of } \pi, \text{residualised features, peer graph}\bigr)
		\]
		denote all quantities fixed before the outcome vector is analysed. Suppose that:
		\begin{enumerate}
			\item Under the fair null, conditional on $\mathcal{G}$, the outcomes are independent and satisfy
			\[
			O_i \mid \mathcal{G} \sim \mathrm{Bern}(p_{0,i}),
			\qquad i=1,\ldots,n,
			\]
			where each $p_{0,i}$ depends only on $(f, s)$.
			
			\item The posterior draws entering $\mathcal{G}$ are computed using $(A, s)$ and $(f, s)$ only, and do not use the observed outcome vector $O$.
			
			\item The probabilities used to generate the Monte Carlo replicates equal the true fair-null probabilities:
			\[
			\widehat{p}_{0,i} = p_{0,i},
			\qquad i=1,\ldots,n.
			\]
		\end{enumerate}
		
		For $b=1,\ldots,B$, independently generate
		\[
		O_b^\star \sim \prod_{i=1}^n \mathrm{Bern}(\widehat{p}_{0,i}),
		\]
		and define
		\[
		p_{\mathrm{cond}}
		=
		\frac{
			1 + \sum_{b=1}^{B} \mathbf{1}\!\left\{ T_{\mathrm{dec}}(O_b^\star) \geq T_{\mathrm{dec}}(O) \right\}
		}{
			B + 1
		}.
		\]
		Then, for every $\varepsilon \in (0,1)$,
		\[
		\Pr\!\left( p_{\mathrm{cond}} \leq \varepsilon \mid \mathcal{G} \right) \leq \varepsilon.
		\]
		Consequently,
		\[
		\Pr\!\left( p_{\mathrm{cond}} \leq \varepsilon \right) \leq \varepsilon.
		\]
		Thus, conditional on the design and nuisance draws, $p_{\mathrm{cond}}$ is a valid, possibly conservative, Monte Carlo $p$-value.
	\end{lemma}
	
	\begin{proof}
		Fix $\mathcal{G}$. After conditioning on $\mathcal{G}$, the observed covariates, protected context, network, posterior draws, residualised features, and peer graph are all treated as fixed. In particular, the decision statistic $T_{\mathrm{dec}}(\cdot)$ is now a fixed function of an outcome vector.
		
		By assumption~(ii), $\mathcal{G}$ was constructed without using $O$. Therefore, conditioning on $\mathcal{G}$ does not condition on any information extracted from the outcomes. Assumption~(i) consequently gives
		\[
		O \mid \mathcal{G} \sim \prod_{i=1}^{n} \mathrm{Bern}(p_{0,i}).
		\]
		
		Next, assumption~(iii) implies that the simulation probabilities equal these same fair-null probabilities:
		\[
		\widehat{p}_{0,i} = p_{0,i}.
		\]
		Hence each simulated outcome vector satisfies
		\[
		O_b^\star \mid \mathcal{G} \sim \prod_{i=1}^{n} \mathrm{Bern}(p_{0,i}).
		\]
		
		The simulated vectors are generated independently of one another and independently of the observed vector. Therefore,
		\[
		(O, O_1^\star, \ldots, O_B^\star) \mid \mathcal{G}
		\]
		are independent and identically distributed.
		
		Now define $V_0 = T_{\mathrm{dec}}(O)$ and, for $b=1,\ldots,B$, $V_b = T_{\mathrm{dec}}(O_b^\star)$. Because $T_{\mathrm{dec}}(\cdot)$ is fixed after conditioning on $\mathcal{G}$, applying it to the conditionally i.i.d.\ outcome vectors preserves exchangeability. Thus,
		\[
		(V_0, V_1, \ldots, V_B) \mid \mathcal{G}
		\]
		are exchangeable.
		
		Define the upper rank of the observed statistic by
		\[
		R = \#\{b \in \{0, 1, \ldots, B\} : V_b \geq V_0\}.
		\]
		This counts the number of statistics that are at least as large as the observed statistic, including $V_0$ itself. Equivalently, $R$ is the rank of $V_0$ when the values are ordered from largest to smallest and ties are counted against the observed statistic.
		
		Since the $B+1$ values are exchangeable, no labelled value is more likely than another to occupy one of the $k$ most extreme positions. Therefore, for every $k=1,\ldots,B+1$,
		\[
		\Pr(R \leq k \mid \mathcal{G}) \leq \frac{k}{B+1}.
		\]
		The inequality may be strict when ties occur, because counting all ties as at least as extreme as $V_0$ makes the test conservative.
		
		Finally, the definition of $R$ gives $p_{\mathrm{cond}} = \frac{R}{B+1}$.
		Hence,
		\[
		\Pr(p_{\mathrm{cond}} \leq \varepsilon \mid \mathcal{G})
		=
		\Pr\!\left( R \leq \lfloor \varepsilon(B+1) \rfloor \mid \mathcal{G} \right).
		\]
		Applying the rank bound with $k = \lfloor \varepsilon(B+1) \rfloor$ yields
		\[
		\Pr(p_{\mathrm{cond}} \leq \varepsilon \mid \mathcal{G})
		\leq
		\frac{\lfloor \varepsilon(B+1) \rfloor}{B+1}
		\leq
		\varepsilon.
		\]
		
		This proves conditional validity. The unconditional result follows by taking expectations over the distribution of $\mathcal{G}$:
		\[
		\Pr(p_{\mathrm{cond}} \leq \varepsilon)
		=
		\mathbb{E}\!\left[ \Pr(p_{\mathrm{cond}} \leq \varepsilon \mid \mathcal{G}) \right]
		\leq
		\varepsilon.
		\]
	\end{proof}

	\section{The data-generating process and the study design}
	\label{app:dgp}\label{sec:sim-dgp}
	Every scenario in the study is a point in a single parameter space. Each simulated dataset contains $n=800$ individuals, with locations $s_1,\dots,s_n$ drawn independently and uniformly on the spatial domain $\mathcal{S}=[0,1]^2$. Figure~\ref{fig:city} shows one simulated city.
	
	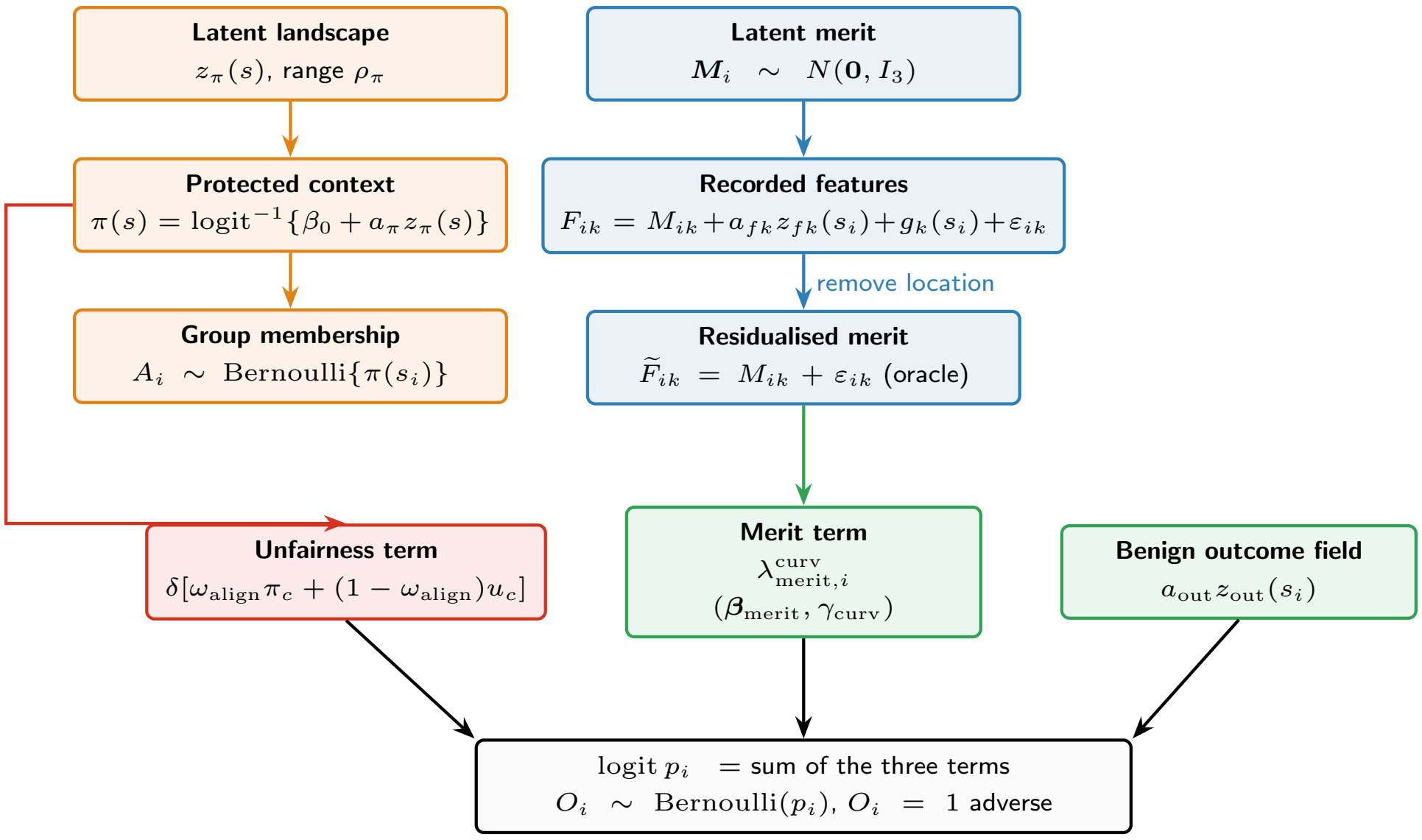
\begin{figure*}[tbp]
		\centering
		\begin{tikzpicture}[
			font=\footnotesize,
			nd/.style={draw,rounded corners=2pt,align=center,inner sep=4pt,
				text width=3.6cm,line width=0.8pt},
			arr/.style={-{Stealth[length=2mm]},thick},
			node distance=5mm and 7mm]
			\node[nd,draw=cProt,fill=cProt!10] (land)
			{\textbf{Latent landscape}\\ $z_\pi(s)$, range $\rho_\pi$};
			\node[nd,draw=cProt,fill=cProt!10,below=of land] (prot)
			{\textbf{Protected context}\\ $\pi(s)=\logit^{-1}\{\beta_0+a_\pi z_\pi(s)\}$};
			\node[nd,draw=cProt,fill=cProt!10,below=of prot] (grp)
			{\textbf{Group membership}\\ $A_i\sim\Bern\{\pi(s_i)\}$};
			\node[nd,draw=cMerit,fill=cMerit!10,right=of land] (mer)
			{\textbf{Latent merit}\\ $\bm M_i\sim N(\bm 0,I_3)$};
			\node[nd,draw=cMerit,fill=cMerit!10,below=of mer,text width=4.4cm] (feat)
			{\textbf{Recorded features}\\ $F_{ik}=M_{ik}+a_{fk}z_{fk}(s_i)+g_k(s_i)+\varepsilon_{ik}$};
			\node[nd,draw=cMerit,fill=cMerit!10,below=of feat] (res)
			{\textbf{Residualised merit}\\ $\Ftil_{ik}=M_{ik}+\varepsilon_{ik}$ (oracle)};
			\node[nd,draw=cFair,fill=cFair!10,below=9mm of res,text width=2.9cm] (mt)
			{\textbf{Merit term}\\ $\lambda^{\mathrm{curv}}_{\mathrm{merit},i}$\\
				$(\bmerit,\gcurv)$};
			\node[nd,draw=cFair,fill=cFair!10,right=of mt,text width=2.9cm] (bo)
			{\textbf{Benign outcome field}\\ $\aout\zout(s_i)$};
			\node[nd,draw=cUnfair,fill=cUnfair!10,left=of mt,text width=3.3cm] (ut)
			{\textbf{Unfairness term}\\ $\dd[\kalign\pic+(1-\kalign)\uc]$};
			\node[nd,draw=black,fill=gray!10,below=9mm of mt,text width=5.6cm] (out)
			{$\logit p_i=$ sum of the three terms\\
				$O_i\sim\Bern(p_i)$, $O_i=1$ adverse};
			\draw[arr,cProt] (land)--(prot);
			\draw[arr,cProt] (prot)--(grp);
			\draw[arr,cMerit] (mer)--(feat);
			\draw[arr,cMerit] (feat)--node[right,font=\scriptsize]{remove location}(res);
			\draw[arr,cFair] (res)--(mt);
			\draw[arr,cUnfair] (prot.west) -| ++(-6mm,0) |- (ut.north);
			\draw[arr] (mt)--(out);
			\draw[arr] (bo.south)--(out.north east);
			\draw[arr] (ut.south)--(out.north west);
		\end{tikzpicture}
		\caption{Anatomy of the data-generating process. Orange: protected context;  blue: merit and its spatial contamination; green: components that are fair with respect to the target estimand; red: the injected unfairness. Each
			green or red component has its own switch ($\gcurv$, $\aout$, $\dd$,
			$\kalign$), so that scenarios differ in exactly the mechanisms they are
			designed to test.}
		\label{fig:anatomy}
	\end{figure*}
	\paragraph{Spatial fields.}
	All latent spatial fields are independent, stationary, isotropic, zero-mean Gaussian random fields with Matérn correlation
	\begin{equation}
		\Cor\{z(s),z(s')\}
		=\frac{2^{1-\nu}}{\Gamma(\nu)}\bigl(\kappa\lVert s-s'\rVert\bigr)^{\nu}
		K_\nu\bigl(\kappa\lVert s-s'\rVert\bigr),
		\qquad \kappa=\frac{\sqrt{8\nu}}{\rho},
		\label{eq:matern}
	\end{equation}
	where $K_\nu$ is the modified Bessel function of the second kind, $\nu$ is the smoothness parameter and $\rho$ is the practical range. Throughout, $\nu=1$. Under this parameterisation, the correlation between two locations separated by a distance $\rho$ is approximately $0.14$. The practical range is therefore the distance beyond which values of the field are effectively unrelated; a larger $\rho$ produces larger, smoother spatial patterns. Each realised field is standardised to unit empirical variance over the $n$ locations before being multiplied by its amplitude, so that amplitudes are comparable across fields and replications. The special case $\rho=0$ denotes spatially unstructured $N(0,1)$ heterogeneity.
	
	\paragraph{Protected context.}
	Let $z_\pi$ be a Matérn field with practical range $\rho_\pi$. The protected
	context is
	\begin{equation}
		\pi(s_i)=\logit^{-1}\{\beta_0+a_\pi z_\pi(s_i)\},\qquad a_\pi=0.90.
		\label{eq:protected}
	\end{equation}
	The amplitude $a_\pi$ governs the degree of residential segregation: when
	$a_\pi=0$, $\pi(s)\equiv\logit^{-1}(\beta_0)$ is constant and group membership carries no spatial structure; as $a_\pi$ increases, the protected group becomes increasingly concentrated in particular areas. The range $\rho_\pi$ governs the spatial scale of that concentration. The
	intercept $\beta_0$ is determined by numerical root finding so that the
	realised mean protected context equals a prespecified minority share,
	\begin{equation}
		\frac1n\sum_{i=1}^{n}\pi(s_i)=\pitarget=0.35.
		\label{eq:rootfind}
	\end{equation}
	The observed protected attribute is then generated as
	$A_i\mid\pi(s_i)\sim\Bern\{\pi(s_i)\}$, independently across individuals.
	Thus $\pi(s_i)$ describes the demographic composition of the setting in
	which individual $i$ lives, and $A_i$ is that individual's realised group
	membership.
	
	\paragraph{Latent merit and recorded features.}
	Each individual has a latent merit feature vector
	$\bm M_i=(M_{i1},M_{i2},M_{i3})^{\top}$, with
	$\bm M_i\sim N(\bm 0,I_3)$ independently of location and of the protected
	context. The recorded merit features are
	\begin{equation}
		F_{ik}=M_{ik}+a_{fk}z_{fk}(s_i)+g_k(s_i)+\varepsilon_{ik},\qquad k=1,2,3,
		\label{eq:features}
	\end{equation}
	where $z_{f1},z_{f2},z_{f3}$ are independent Matérn fields with common
	practical range $\rho_f$, the amplitudes are
	$(a_{f1},a_{f2},a_{f3})=(0.70,0.50,0.35)$, and
	$\varepsilon_{ik}\sim N(0,\sigma_f^2)$ with $\sigma_f=0.20$. The terms
	$a_{fk}z_{fk}(s)+g_k(s)$ represent location-associated advantage in the
	recorded features. The deterministic trends are
	
	\begin{align}
		g_1(s) &= \sin(2\pi x) + \frac{1}{2}\cos(3\pi y) \\
		g_2(s) &= 3\left(x - \frac{1}{2}\right)^2 - 2\left(y - \frac{1}{2}\right)^2 \\
		g_3(s) &= 0.8\sin\left\{\pi(x + y)\right\}
	\end{align}
	with $s=(x,y)$; because they are not of the form assumed by the
	residualisation model of Section~\ref{sec:B-Fair}, the feasible analysis of Part~II	cannot recover the location effect exactly. In the oracle analysis of
	Part~I the residualised features are supplied at their true values,
	\begin{equation}
		\Ftil^{\,\mathrm{oracle}}_{ik}=F_{ik}-a_{fk}z_{fk}(s_i)-g_k(s_i)
		=M_{ik}+\varepsilon_{ik}.
		\label{eq:oracle-resid}
	\end{equation}
	
	
	\paragraph{Outcome mechanism.}
	The outcome $O_i\in\{0,1\}$ is coded so that $O_i=1$ is the adverse
	outcome. Its log-odds combine four components. The merit component is
	$\lambda_{\mathrm{merit},i}=\Fbtil_i^{\top}\bmerit$ with fixed coefficients
	$\bmerit=(0.9,-0.6,0.4)^{\top}$. To permit functional-form misspecification
	of the merit--outcome relationship, it is extended to
	\begin{equation}
		\lambda^{\mathrm{curv}}_{\mathrm{merit},i}
		=\lambda_{\mathrm{merit},i}
		+\gcurv\Bigl(\lambda^{2}_{\mathrm{merit},i}
		-\frac1n\sum_{j=1}^{n}\lambda^{2}_{\mathrm{merit},j}\Bigr),
		\label{eq:curv}
	\end{equation}
	where centring gives the curvature term, sample mean zero on the
	linear-predictor scale, so that $\gcurv$ changes the shape of the
	relationship without shifting its level. The misspecification scenarios use
	$\gcurv=0.35$. Benign spatial outcome dependence enters through
	$\aout\zout(s_i)$, where $\zout$ is a Matérn field with practical range
	$\rhoout$, independent of the protected context and of latent merit by
	construction.
	
	The unfairness component is defined relative to the centred protected
	context $\pic(s_i)=\pi(s_i)-\pibar$, where
	$\pibar=n^{-1}\sum_j\pi(s_j)$. Let $u$ be a further independent Matérn field
	with practical range $\rho_\pi$, and let $\uc$ denote $u$ centred and
	rescaled to the realised empirical standard deviation of $\pic$,
	\begin{equation}
		\uc(s_i)=\frac{u(s_i)-\bar u}{\sd\{u(s_1),\dots,u(s_n)\}}\,
		\sd\{\pic(s_1),\dots,\pic(s_n)\}.
		\label{eq:uc}
	\end{equation}
	The outcome model is then
	\begin{equation}
		\begin{aligned}
			\logit(p_i)&=\lambda^{\mathrm{curv}}_{\mathrm{merit},i}+\aout\zout(s_i)\\
			&\quad+\dd\bigl\{\kalign\pic(s_i)+(1-\kalign)\uc(s_i)\bigr\},\\
			O_i\mid p_i&\sim\Bern(p_i).
		\end{aligned}
		\label{eq:outcome}
	\end{equation}
	There is no outcome intercept, so the mean prevalence is close to
	one-half.
	
	The parameter $\dd\ge 0$ sets the magnitude of the injected unfairness, and
	$\kalign\in[0,1]$ sets its alignment with the protected context: at
	$\kalign=1$ the disadvantage tracks $\pic$ exactly, and at $\kalign=0$ it is
	spatially structured but unrelated to the protected context. Because $\pic$
	and $\uc$ are centred and share a common empirical spread, changing $\dd$ or
	$\kalign$ does not shift the mean of the linear predictor. Since the inverse-logit link is nonlinear, realised prevalence may nevertheless vary modestly across
	scenarios.
	
	A mechanism is \emph{substantively fair} with respect to the target estimand
	if and only if $\dd=0$. This holds whatever the values of $\aout$ and
	$\gcurv$: a fair city may contain spatially clustered outcomes and a
	nonlinear merit--outcome relationship. Unless otherwise stated,
	$\rho_\pi=\rho_f=0.25$, $\pitarget=0.35$, $\kalign=1$ and $\gcurv=0$; whenever benign spatial outcome dependence is present, $\aout=0.80$ and
	$\rhoout=0.30$. These stress values are chosen so that the benign field varies on a spatial scale comparable to the protected context ($0.30$ against $0.25$) and is large on the log-odds scale relative to the injected unfairness term (Figure~\ref{fig:city}); this is the configuration in which a fair-null model can least easily separate the two, and so the most demanding test of both validity and sensitivity.
	
	
	\subsection{Structure of the study}
	\label{sec:structure}\label{sec:part1-decision-mechanisms}
	
	The study has two parts, which differ in what the audit is permitted to
	know.
	
	\textbf{Part I (oracle analysis)} supplies the protected context $\pi(s)$
	and the residualised features $\Fbtil$ at their true simulated values. This
	removes all estimation error from the nuisance quantities, so that any
	failure of validity or sensitivity can be attributed to the directional
	functional $\That$ itself or to the fair-null reference distribution against
	which it is judged. Part~I therefore assesses the mathematical and
	inferential structure of \BFair{}.
	
	\textbf{Part II (feasible Bayesian analysis)} requires the audit to estimate
	$\pi(s)$ and $\Fbtil$ from the data by SPDE--INLA \citep{lindgren2011explicit,rue2009inla} and to propagate posterior
	uncertainty through $\That$, as it would in an applied audit. Part~II
	therefore assesses how the procedure behaves under the conditions in which
	it would actually be deployed.
	
	The principal scenarios form a matched $3\times2$ design (Table~\ref{tab:design}). Rows are three mechanisms that differ in exactly one switch at a time; columns are the two parts. Within a row the data-generating process is identical, so a difference between columns is attributable to estimating the nuisance quantities. Moving down a row changes first the benign outcome dependence $\aout$ and then the unfairness magnitude $\dd$, so that the rows isolate the transport effect and the sensitivity effect in turn. In the third row the feasible cell is run on the same simulated datasets as the oracle cell, dataset by dataset, so the comparison between parts is paired.
	
	A further oracle family (R7) varies the lattice side length of the restricted-spatial reference, $r\in\{2,\dots,6\}$, with a fair arm ($\dd=0$) and an unfair arm ($\dd=1.25$) at $\aout=0.80$, 500 datasets per arm. Rule~D1, fixed before the R7 results were examined, adopts the smallest lattice whose level is not rejected at the $0.10$ significance level and whose sensitivity is within five percentage points of the most sensitive lattice.
	
	\paragraph{Part I audits}
	With $\pi$ and $\Fbtil$ known, each dataset yields one value of $\That$, and the audit rejects when the plus-one Monte Carlo $p$-value over $B=499$ outcome resamples is at most $0.05$. The resamples regenerate only $O$, holding $\Dfixed$ and the peer graph fixed, and use common random numbers across references, so contrasts between references are paired. Three references are reported (Table~\ref{tab:reference-distributions}).
	
	\paragraph{Part II audits}
	Here $\pi(s)$ is estimated by a binomial SPDE model for $A$ and each feature's location effect by a Gaussian SPDE model \citep{lindgren2011explicit}, on one mesh per dataset, fitted by INLA \citep{rue2009inla} with penalised-complexity priors \citep{simpson2017penalising,fuglstad2019constructing} satisfying $\Pr(\rho<0.1)=0.05$ and $\Pr(\sigma>1.5)=0.05$. Hyperparameters are fixed at their posterior mode, and $M=1000$ joint draws are paired across the two independently fitted models. The \emph{plugin} variant builds the peer graph once from the posterior-mean features; the \emph{full} variant rebuilds it at every draw; the \emph{known-nuisance benchmark} substitutes the true $\pi$ and $\Fbtil$, so its posterior is degenerate at the oracle $\That$. Every variant is decided under a transported threshold; the benchmark and plugin variants are also decided by design-conditional calibration.
	\begin{itemize}
		\item \emph{Transported threshold.} $\tau_v$ is the empirical $0.95$ quantile of $T_{\mathrm{dec},v}$ over $\Bcal=2000$ B1 datasets, so that $5\%$ of fair B1 datasets exceed it. This is a quantile of the sampling distribution of $\Tdec$ across fair populations, not of the within-population reference distribution $\mathcal R_0$; the posterior tail level $\alpha=0.05$ of Equation~\eqref{eq:rule-prob} and the $0.95$ quantile used here play different roles. The B1 datasets are drawn on a random-number stream disjoint from every evaluation stream, and the threshold is then held fixed for all cells. The thresholds are $\tau_{\mathrm{benchmark}}=0.04832$, $\tau_{\mathrm{plugin}}=0.04407$ and $\tau_{\mathrm{full}}=0.04393$ (bootstrap standard errors $0.00028$, $0.00022$, $0.00021$). Because the benchmark's posterior is degenerate, $\tau_{\mathrm{benchmark}}$ is an independent estimate of the 0.95 quantile of the oracle statistic under B1; it agrees with the oracle-layer quantile $0.04841$ estimated from the pooled P1 ensemble, as it should.
		\item \emph{Design-conditional.} $p_{\mathrm{cond}}$ of Equation~\eqref{eq:pcond} with $B=499$ resamples from the restricted-spatial fair null, on the B3 and B16 datasets. Rule~D1 adopted no lattice (Section~\ref{sec:sim-results}), so the primary lattice was fixed before any design-conditional result was available as $r=3$, the smallest lattice whose oracle-layer level was not rejected; $r=2$, the most sensitive lattice, is reported as secondary.
	\end{itemize}
	
	\paragraph{Monte Carlo error}
	Rates under a transported threshold are reported with $95\%$ Wilson intervals at the effective size $\Beff=B/(1+B/\Bcal)$, which accounts for the error in the estimated threshold; with $B=500$, $\Beff=400$. Rates obtained from per-dataset $p$-values, in Part~I and in the design-conditional audit, use plain Wilson intervals. Paired comparisons use the exact McNemar test, and all reported McNemar $p$-values are two-sided.

	\section{Further results}
	\begin{table*}[htbp]
		\centering
		\caption{Distribution of the decision statistic across datasets under B1 (where $\tau_v$ is calibrated) and B3, and the B3 false-rejection rate decomposed under a Gaussian approximation. ``Location only'' keeps the B1 spread and moves the mean to its B3 value; ``spread only'' does the reverse.}
		\label{tab:locscale}
		\footnotesize\setlength{\tabcolsep}{3pt}
		\begin{tabular}{lcccccccc}
			\toprule
			& \multicolumn{2}{c}{B1} & \multicolumn{2}{c}{B3} & & \multicolumn{3}{c}{B3 rejection (\%), Gaussian}\\
			Variant & mean & s.d. & mean & s.d. & B3 observed (\%) & location only & spread only & both\\
			\midrule
			benchmark & 0.04199 & 0.00342 & 0.04357 & 0.00564 & 21.0 & 8.3 & 13.1 & 20.0\\
			plugin    & 0.03531 & 0.00503 & 0.03555 & 0.00606 &  9.2 & 4.5 &  7.4 &  8.0\\
			full      & 0.03558 & 0.00500 & 0.03578 & 0.00601 &  9.6 & 5.1 &  8.2 &  8.7\\
			\bottomrule
		\end{tabular}
	\end{table*}
	\begin{table*}[htbp]
		\centering
		\caption{Law-of-total-variance decomposition Equation~\eqref{eq:ltv} of $\That$ in Part~I, using the exact pedestal $\theta_0(\mathcal D)$ recorded for every dataset. The cross term is below $1.2\times10^{-6}$ in every row.}
		\label{tab:ltv}
		\footnotesize
		\begin{tabular}{lccccc}
			\toprule
			Cell & $B$ & $\overline{\theta}_0$ & s.d.$(\That)$ & $[\E\{\Var(\That\mid\mathcal D)\}]^{1/2}$ & s.d.$\{\theta_0(\mathcal D)\}$ (share)\\
			\midrule
			P1 (fair, $\aout=0$)     & 1500 & 0.0424 & 0.0037 & 0.0035 & 0.0010 (0.08)\\
			P3 (fair, $\aout=0.8$)   & 1000 & 0.0439 & 0.0057 & 0.0034 & 0.0046 (0.67)\\
			P4 (unfair, $\aout=0.8$) &  500 & 0.0517 & 0.0063 & 0.0034 & 0.0050 (0.65)\\
			\bottomrule
		\end{tabular}
	\end{table*}
	\label{app:results}
	\paragraph{Location and spread of the decision statistic.}
	Table~\ref{tab:locscale} reports the distribution of $\Tdec$ across fair cities with and without benign clustering. Under a Gaussian approximation, the transported threshold over-rejects exactly when $\mu_1-\mu_0>z_{0.95}(\sigma_0-\sigma_1)$, where $(\mu_0,\sigma_0)$ and $(\mu_1,\sigma_1)$ are the mean and standard deviation of $\Tdec$ where the threshold was calibrated and where it is applied; with the mean unchanged, any increase in spread is enough. The approximation reproduces the observed rates to within 1.2 percentage points, and the spread inflation accounts for most of the excess.
	
	\paragraph{Variance decomposition.}
	Across cities,
	\begin{equation}
		\Var(\That)=\E\{\Var(\That\mid\mathcal D)\}+\Var\{\theta_0(\mathcal D)\}.
		\label{eq:ltv}
	\end{equation} Table~\ref{tab:ltv} evaluates both terms in the oracle analysis, where the pedestal is known for every city.
	
	\paragraph{Paired comparisons.}
	On identical fair cities with benign clustering, 78 are rejected by the known-nuisance benchmark alone against 19 by the feasible plugin audit alone under the transported threshold (exact McNemar $p=1.2\times10^{-9}$); adding curvature to the merit--outcome relationship changes neither conclusion (benchmark 18.0\%, plugin 8.6\%). Design-conditional calibration against transported thresholds on the same cities: benchmark 11 conditional-only against 85 transported-only rejections ($p=2.5\times10^{-15}$); plugin 13 against 42 ($p=1.1\times10^{-4}$). At $r=2$ the conditional false-rejection rates are 9.4\% (benchmark) and 7.2\% (plugin), and the contrast with $r=3$ is significant (plugin 25 against 6, $p=8.8\times10^{-4}$). In unfair cities, the conditional and transported plugin audits disagree on 64 against 89 cities ($p=0.052$).
	
	\paragraph{Propagation and priors.}
	Rebuilding the peer graph at every posterior draw changed few decisions: the plugin and full variants disagreed on 10 of 498 (B1), 8 of 500 (B3) and 13 of 500 (B16) cities, while widening the posterior of $\Tdir$ by 6.4--9.6\% in standard deviation. Changing the penalised-complexity prior medians did not change decisions detectably under the fair baseline (all paired McNemar $p\ge0.125$).
\end{document}